\documentclass[aps,twocolumn,prc,showpacs]{revtex4}
\usepackage{mathrsfs}
\usepackage{longtable,lscape}
\usepackage{txfonts}
\usepackage{amssymb}
\usepackage{indentfirst}
\usepackage{graphicx,,booktabs}
\usepackage{color}
\usepackage{amssymb}
\usepackage{epsfig}

\newcommand{\vsig}{\mbox{\boldmath$\sigma$\unboldmath}}

\newcommand{\be}{\begin{equation}}
\newcommand{\ee}{\end{equation}}
\newcommand{\bea}{\begin{eqnarray}}
\newcommand{\eea}{\end{eqnarray}}
\newcommand{\bean}{\begin{eqnarray*}}
\newcommand{\eean}{\end{eqnarray*}}

\newcommand{\gapproxeq}{\lower
.7ex\hbox{$\;\stackrel{\textstyle >}{\sim}\;$}}
\newcommand{\lapproxeq}{\lower
.7ex\hbox{$\;\stackrel{\textstyle <}{\sim}\;$}}

\begin{document}

\title{Quark model study of the $\pi N\to \pi N$ reactions up to the $N(1440)$ resonance region }
\author{
Kai-Lei Wang, Li-Ye Xiao, Xian-Hui Zhong~\footnote {E-mail: zhongxh@hunnu.edu.cn}
} \affiliation{ 1) Department
of Physics, Hunan Normal University, and Key Laboratory of
Low-Dimensional Quantum Structures and Quantum Control of Ministry
of Education, Changsha 410081, China }

\affiliation{ 2) Synergetic Innovation
Center for Quantum Effects and Applications (SICQEA),
Hunan Normal University,Changsha 410081,China}

\begin{abstract}
A combined analysis of the reactions $\pi^+p\to \pi^+p$,
$\pi^-p\to \pi^-p$ and $\pi^-p\to \pi^0n$ is carried
out with a chiral quark model. The observations
are reasonably described from the $\Delta(1232)$ resonance region up to the $N(1440)$ resonance region.
Besides the $\Delta(1232)P_{33}$, a confirmed role of $N(1440)P_{11}$
is found in the polarizations of
the $\pi^-p\to \pi^-p$ and $\pi^-p\to\pi^0n$ reactions. It is found that the
$N(1440)N\pi$ and $\Delta(1232)N\pi$ couplings are about $1.7$ and $4.8$
times larger than the expectations from the simple quark model, respectively,
which may suggest the unusual property of $N(1440)P_{11}$ and deficiency of
the simple quark model in the description of $N(1440)P_{11}$ and $\Delta(1232)P_{33}$.
The $t$- and $u$-channel backgrounds have notable
contributions to the $\pi^+ p\rightarrow\pi^+ p$ reaction, while
in the $\pi^-p\rightarrow \pi^- p,\pi^0 n$ reactions,
the $s$-channel nucleon and $t$- and $u$-channel backgrounds
play crucial roles.
\end{abstract}

\maketitle

\section{Introduction}{\label{introduction}}

A better understanding of the baryon spectrum and internal structure of excited baryons is a fundamental
challenge and goal in hadronic physics~\cite{Klempt:2009pi,Crede:2013sze,Briscoe:2015qia}.
Pion-nucleon ($\pi N$) scattering provides us an important place to study the $\Delta$
and nucleon spectroscopies. Most of our current knowledge about the $\Delta$
and nucleon resonances listed in the Review of Particle Physics by the Partial Data Group (PDG)~\cite{PDG}
was extracted from the $\pi N$ scattering. In the past decades, although many efforts have
been made by several partial wave analysis groups~\cite{Wu:2016ixr,Koch:1980ay,Cutkosky:1979zv,Cutkosky:1979fy,Cutkosky:1990zh,
Arndt:1985vj,Arndt:2006bf,Krehl:1999km,Doring:2009yv,Kamano:2010ud,Ronchen:2012eg,Kamano:2013iva,
Suzuki:2009nj,Matsuyama:2006rp,JuliaDiaz:2007kz,Chen:2007cy,Penner:2002ma,
Shklyar:2004ba,Anisovich:2004zz,Anisovich:2010an,Ceci:2006ra,Svarc:2014aga,Svarc:2014zja}, the properties of some $\Delta$ and
nucleon resonances are not well understood. Still, strong model dependencies exist
in the extracted resonance properties from different groups.
For example, the study of $\pi N$ scattering in the literature~\cite{Krehl:1999km,Doring:2009yv}
indicates that the Roper $N(1440)P_{11}$ is dynamically generated
from the coupled channel interaction without any excited three-quark core,
while in the literature~\cite{Chen:2007cy,Suzuki:2009nj,Segovia:2015hra} the Roper $N(1440)P_{11}$ is suggested to
be a three-quark state dressed by a meson cloud.
Furthermore, in some literature the $N(1535)S_{11}$ resonance is suggested to be
a dynamically generated resonance by analyzing the $\pi N$ reactions~\cite{Doring:2008sv,Doring:2009uc,
Kaiser:1995cy,Nieves:2001wt,Inoue:2001ip}. Recently, according to our chiral
quark model study of the $\pi^-p\to \eta n,K^+\Lambda$~\cite{Xiao:2016dlf,Zhong:2007fx}, and $\gamma N\to \eta N,\pi^0 N$ reactions~\cite{Xiao:2015gra,Zhong:2011ti},
the $N(1535)S_{11}$ resonance can be explained as a
mixing three-quark state between representations of $[70,^28]$ and
$[70,^48]$. To deepen our understanding of the resonance
properties from the $\pi N$ reactions, more partial wave analyses
are needed.

In present work, we further extend the chiral quark model to the study
of the $\pi^{\pm}$ elastic reactions $\pi^+p\rightarrow \pi^+p$,
$\pi^-p\rightarrow \pi^-p$ and the charge-exchange reaction $\pi^-p\rightarrow \pi^0n$
up to the $N(1440)$ resonance region.
The $\pi N\to \pi N$ reactions provide us a good place to study
the $\Delta(1232)P_{33}$ and $N(1440)P_{11}$, because the other higher resonances, such as
$N(1535)S_{11}$ and $\Delta(1620)S_{31}$ are far from
the $\Delta(1232)$ and $N(1440)$ region,
their interferences in this low energy region should be strongly suppressed
by the phase space. On the other hand, in the energy regions what we will consider,
there are abundant data, which have been collected by the GWU group~\cite{INS:Data}.
By a combined analysis of these reactions,
we hope (i) to further test the validity of the chiral quark model and
obtain a better understanding of the reaction mechanism for the $\pi N$ scattering;
(ii) to confirm the properties of $\Delta(1232)P_{33}$ extracted from the
$\pi^0$-meson photoproduction processes in our previous work~\cite{Xiao:2015gra}; (iii)
to extract some reliable information of $N(1440)P_{11}$. In our previous quark model analyses of the
$\pi^-p\rightarrow \eta n,K^+\Lambda$~\cite{Xiao:2016dlf,Zhong:2007fx}
and $\gamma N\to \eta N, \pi^0 N$~\cite{Xiao:2015gra,Zhong:2011ti} reactions, no
obvious evidence of $N(1440)P_{11}$ is found.

In the chiral quark model, an effective chiral Lagrangian is
introduced to account for the quark-pseudoscalar-meson coupling.
Since the quark-meson coupling is invariant under the chiral
transformation, some of the low-energy properties of QCD are
retained. There are several outstanding features for this model
~\cite{ Zhong:2007fx,Li:1997gd,Zhao:2010jc}. One is that in this framework
only one overall parameter is needed for the nucleon resonances to
be coupled to the pseudoscalar mesons in the SU(6)$\otimes$O(3)
symmetry limit. This is distinguished from hadronic models where each
resonance requires one additional coupling constant as free
parameter. Furthermore, the $s$- and $u$-channel transition amplitudes at the
tree level can be explicitly calculated, and the quark
model wavefunctions for the baryon resonances, after convolution integrals,
provide a form factor for the interaction vertices. Consequently, all the baryon resonances
can be consistently included.
The chiral quark model has been well developed and successfully applied
to pseudoscalar-meson photoproduction
reactions~\cite{Xiao:2015gra,Zhong:2011ti,Li:1997gd,Zhong:2011ht,Li:1994cy,Li:1995si,
Li:1995vi,Zhao:2002id,Li:1998ni,Zhao:2010jc,Saghai:2001yd,Zhao:2000iz,He:2008ty,He:2008uf}. Recently, this model has been extended to
$\pi^- p$~\cite{Xiao:2016dlf,Zhong:2007fx} and $K^-p$~\cite{Xiao:2013hca,Zhong:2008km,Zhong:2013oqa} reactions as well,
which provides some novel insights into the observables measured in
these reactions.

This work is organized as follows. The model is reviewed in Sec.II.
Then, in Sec.III, our numerical results and analysis are presented
and discussed. Finally, a summary is given in Sec.IV.

\section{Framework}

In this section, we give a brief review of the chiral quark model.
In this model, the meson-quark interactions are adopted by the
effective chiral Lagrangian~\cite{Li:1997gd}
\begin{equation}
H_m=\frac{1}{f_m}\bar{\psi}_j\gamma^j_{\mu}\gamma^j_5\psi_j\vec{\tau}\cdot\partial^{\mu}\vec{\phi}_m,
\end{equation}
where $\psi_j$ represents the $j$-th quark field in a hadron, $f_m$  is
the meson's decay constant, and $\phi_m$ is the field of the
pseudoscalar-meson octet. Then the $s$- and $u$-channel transition amplitudes $\mathbf{\cal M}_s$
and $\mathbf{\cal M}_u$ can be worked out with the relations~\cite{Zhong:2007fx}:
\begin{eqnarray}\label{ms}
\mathbf{\cal M}_s=\sum_j\langle N_f|H_m^f|N_j\rangle\langle N_j|\frac{1}{E_i+\omega_i-E_j}H_m^i|N_i\rangle,\\
\mathbf{\cal M}_u=\sum_j\langle N_f|H_m^i\frac{1}{E_i-\omega_i-E_j}|N_j\rangle\langle N_j|H_m^f|N_i\rangle.
\end{eqnarray}
In the above equations, the $\omega_i$ and $\omega_f$ are the
energies of the incoming and outgoing mesons, respectively.
$|N_i\rangle$, $|N_j\rangle$ and $|N_f\rangle$ stand for the
initial, intermediate, and final states, respectively, and their
corresponding energies $E_i$, $E_j$, and $E_f$ are the
eigenvalues of the nonrelativistic Hamiltonian of the constituent
quark model $\hat{H}$~\cite{Isgur:1978xj,Isgur:1977ef,Isgur:1978wd}. In our previous work
~\cite{Zhong:2008km, Zhong:2007fx}, the
amplitudes $\mathbf{\cal M}_s$ and $\mathbf{\cal M}_u$ have been worked out in the harmonic
oscillator basis.

The $t$-channel backgrounds might play
an important role in the reactions, thus, the $t$-channel contributions of vector exchange and the scalar
exchange are considered in this work. The vector meson-quark and
scalar meson-quark interactions are adopted by~\cite{Xiao:2016dlf}
\begin{eqnarray}
H_V&=&\bar{\psi}_j\left(a\gamma^{\nu}+\frac{b\sigma^{\nu\lambda}\partial_{\lambda}}{2m_q}\right)V_{\nu}\psi_j,\\
H_S&=&g_{Sqq}\bar{\psi}_j\psi_jS.
\end{eqnarray}
Meanwhile, the $VPP$ and $SPP$ couplings  are adopted as
\begin{eqnarray}
H_{VPP}&=&-iG_VTr([\phi_m,\partial_\mu\phi_m]V^{\mu}),\\
H_{SPP}&=&\frac{g_{SPP}}{2m_\pi}\partial_\mu\phi_m\partial^\mu\phi_m
\end{eqnarray}
where $V$, $P$ and $S$ stand for the vector-, pseudoscalar-, scalar-meson fields, respectively.
The coupling constants $a$, $b$, $g_{Sqq}$, $G_V$, and $g_{SPP}$
are to be determined by experimental data. In this work, both the scalar
$\sigma$- and vector $\rho$-meson exchanges are considered for
the $\pi^+p\to \pi^+p$  and $\pi^-p\to \pi^-p$
processes, while the vector $\rho$-meson exchange is only
considered for the $\pi^-p\rightarrow \pi^0p$ process.
The details of the $t$-channel transition amplitude can be found in our previous work~\cite{Zhong:2013oqa}.

Furthermore, the backgrounds from the Coulomb interactions and the contract term maybe play some
roles in the reactions at low energies. To include the contributions from the contract term (meson-meson-quark-quark interaction), we adopt an
effective chiral Lagrangian~\cite{Lutz:2001yb}:
\begin{equation}
H_{contact}=\frac{i}{4f_m^2}\bar{\psi}_j\gamma^{\mu}[[\phi_{m},(\partial_{\mu}\phi_{m})],\psi_j],
\end{equation}
To include the contributions of Coulomb interactions, we follow the method developed
in Refs.~\cite{Tromborg:1976bi,Gashi:2000es,Gashi:2000et,Matsinos:2006sw}. The details
of the amplitudes for the Coulomb term can be found in Ref.~\cite{Matsinos:2006sw}.

In this work, we focus on the contributions of the $s$-channel resonances, which are
degenerate within the same principle number $n$.
To obtain the contributions of individual resonances, we need to separate out the
single-resonance-excitation amplitudes within each principle number
$n$ in the $s$ channel. Taking into account the width effects of the
resonances, the resonance transition amplitudes of $s$ channel can be
generally expressed as~\cite{Zhong:2007fx,Zhong:2013oqa}
\begin{equation}
\mathcal{M}_R^s=\frac{2M_R}{s-M_R^2+iM_R\Gamma_R}\mathcal{O}_Re^{-(\mathbf{k}^2+\mathbf{q}^2)/6\alpha^2},
\end{equation}
where $\sqrt{s}=E_i+\omega_i$ is the total energy of the system, $\mathbf{k}$
and $\mathbf{q}$ stand for the momenta of incoming and outing mesons, $\alpha$
is the harmonic oscillator strength, and $\mathcal{O}_R$ is the
separated operators for individual resonances.
$M_R$ is the mass of the $s$-channel resonance with a width
$\Gamma_R$. The transition
amplitude can be written in a standard form~\cite{Hamilton:1963zz}:
\begin{equation}
\mathcal{O}_R=f(\theta)+i g(\theta)\vsig\cdot \mathbf{n},
\end{equation}
where $\vsig$ is the spin operator of the nucleon,
$\mathbf{n}\equiv \mathbf{q}\times \mathbf{k}/|\mathbf{k}\times
\mathbf{q}|$. $f(\theta)$ and $g(\theta)$ stand for the
non-spin-flip and spin-flip amplitudes, respectively, which can be
expanded in terms of the familiar partial wave amplitudes $T_{l\pm}$
for the states with $J=l\pm 1/2$:
\begin{eqnarray}\label{f wave}
f(\theta)&=&\sum_{l=0}^\infty[(l+1)T_{l+}+lT_{l-}]P_l(\cos\theta),\\
g(\theta)&=&\sum_{l=0}^\infty[T_{l-}-T_{l+}]\sin\theta P_l^\prime(\cos\theta).
\end{eqnarray}

Both the isospin-$\frac{1}{2}$ and isospin-$\frac{3}{2}$ resonances contribute
to the $\pi^-p\rightarrow \pi^-p,\pi^0n$ reactions.
Thus, we need separate out the isospin-$\frac{1}{2}$ and
$\frac{3}{2}$ resonance contributions from these reaction amplitudes.
As we know, the partial wave amplitudes $T_{l\pm }$ for the
$\pi N\to \pi N $ reactions can be decomposed into
the linear combinations of $s$-channel isospin amplitudes with the relations
\begin{eqnarray}
T_{l\pm }(\pi^+p\to\pi^+p )&=&T^{3/2}_{l\pm },\\
T_{l\pm }(\pi^-p\to\pi^-p )&=&+\frac{1}{3}(2T^{1/2}_{l\pm }+T^{3/2}_{l\pm }),\\
T_{l\pm }(\pi^-p\to\pi^0n )&=&-\frac{\sqrt{2}}{3}(T^{1/2}_{l\pm }-T^{3/2}_{l\pm }),
\end{eqnarray}
where $T^{1/2}$ and $T^{3/2}$ correspond to the isospin-$\frac{1}{2}$, and
$\frac{3}{2}$ resonance contributions, respectively.
Using these relations, we can separate out $s$-channel isospin contributions
from the $\pi N\to \pi N$ amplitudes.

In the SU(6)$\otimes$O(3) symmetry limit, we have extracted the amplitudes for each
$s$-channel resonances within $n\leq 2$ shell for
the $\pi^+p\rightarrow \pi^+p $, $\pi^-p\rightarrow \pi^-p$ and
$\pi^-p\rightarrow \pi^0n$ processes. Our results are listed in
Tables~\ref{amplitudes} and~\ref{amplitudes2}.  Comparing the amplitudes
of different resonances with each other, one can easily find which
states are the main contributors to the reactions in the
SU(6)$\otimes$O(3) symmetry limit.

Finally, the differential cross section $d\sigma/d\Omega$ and polarization $P$ can be calculated by
\begin{eqnarray}\label{weifenjiemian}
\frac{d\sigma}{d\Omega}&=&\frac{(E_i+M_i)(E_f+M_f)}{64\pi^2s(2M_i)(2M_f)}\frac{|\mathbf{q}|}{|\mathbf{k}|}\frac{1}{2}
\times\sum_{\lambda_i,\lambda_f}\left|M_{\lambda_f,\lambda_i}\right|^2,\nonumber
\end{eqnarray}
\begin{eqnarray}\label{jihua}
P=2\frac{\mathrm{Im}[f(\theta)g^{\*}(\theta)]}{|f(\theta)|^2+|g(\theta)|^2},
\end{eqnarray}
where $\lambda_i=\pm\frac{1}{2}$ and $\lambda_f=\pm\frac{1}{2}$ are the helicities of the initial and final state baryons respectively.

\begin{table}
\begin{center}
\caption{Parameters.}\label{par1}
\begin{tabular}{ccccccccc}
\hline\hline
 Constituent quark mass    ~~~&$M_{u}$       &330 MeV\\
                              &$M_{d}$       &330 MeV\\
                              &$M_{s}$       &450 MeV\\
Harmonic oscillator parameter~~~&$\alpha$       &400 MeV\\
\hline
degenerate masses of the ~~~&$M_1$     &1650 MeV\\
$n=1,2$ shell resonances ~~~&$M_2$     &1750 MeV\\
\hline
Parameters in $t$ channel~~~&$G_{V}a$  &$12$    \\
                         ~~~&$g_{SPP}g_{Sqq}$  &$65$ \\
                         ~~~&$m_{\rho}$  &770 MeV \\
                         ~~~&$m_{\sigma}$  &450 MeV \\
\hline
 $\pi NN$ coupling               &$g_{\pi NN}$  &$13.48$\\
\hline\hline
\end{tabular}
\end{center}
\end{table}

\begin{table*}[htb]
\begin{center}
\caption{ \label{compare} Masses $M_{R}$ (MeV) and widths $\Gamma_{R}$ (MeV) of $s$-channel intermediate states, and their $C_R$ strength parameters.}
\footnotesize
\begin{tabular}{|p{1.7cm}|ccc|ccc|ccc|cc|c}
\hline\hline
Resonance             & \multicolumn{3}{|c|} {\underline{$\pi^+p\rightarrow \pi^+p$}}   & \multicolumn{3}{|c|} {\underline{$\pi^-p\rightarrow \pi^-p$}}     & \multicolumn{3}{|c|} {\underline{$\pi^-p\rightarrow \pi^0n$}}         \\
                          &$\Gamma_{R}$&$M_{R}$&$C_{R}$              &$\Gamma_{R}$&$M_{R}$&$C_{R}$                      &$\Gamma_{R}$&$M_{R}$&$C_{R}$ \\
\hline
$N(938)P_{11}$&$\cdot\cdot\cdot$&$\cdot\cdot\cdot$&$\cdot\cdot\cdot$&$\cdot\cdot\cdot$&938&                                                                     $1.0 $&$\cdot\cdot\cdot$&                                    938&$1.0 $           \\
$\Delta(1232)P_{33}$       &$100^{+7}_{-4}$&$1212^{+3}_{-5}$&$2.95^{+0.05}_{-0.20}$    &$100^{+7}_{-3}$&$1212^{+2}_{-5}$&$3.00^{+0.05}_{-0.15}$                &$100^{+5}_{-2}$&$1210^{+2}_{-5}$&$3.10^{+0.02}_{-0.15}$         \\
$N(1440)P_{11}$        &$\cdot\cdot\cdot$&$\cdot\cdot\cdot$&$\cdot\cdot\cdot$                 &$200^{+40}_{-35}$&$1400^{+40}_{-25}$&$23 \pm 6$        &$200_{-20}^{+35}$&$1400^{+40}_{-25}$&$23^{+6}_{-4}$    \\
$N(1535)S_{11}$       &$\cdot\cdot\cdot$&$\cdot\cdot\cdot$&$\cdot\cdot\cdot$                                              &$124 $&$1524 $&$1.0$                             &$124$&$1524$&$1.0$         \\
$\Delta(1620)S_{31}$       &$140$&$1630$&$1.0$                         &$140$&$1630$&$1.0$                               &$140$&$1630$&$1.0$     \\
$N(1650)S_{11}$        &$\cdot\cdot\cdot$&$\cdot\cdot\cdot$&$\cdot\cdot\cdot$                                             &$119$&$1670$&$1.0$                               &$119$&$1670$&$1.0$     \\
$\Delta(1700)D_{33}$    &$300$&$1745$&$1.0$                            &$300$&$1745$&$1.0$                               &$300$&$1745$&$1.0$      \\
$N(1520)D_{13}$         &$\cdot\cdot\cdot$&$\cdot\cdot\cdot$&$\cdot\cdot\cdot$                &$125$&$1515$&$1.0$               &$125$&$1515$&$1.0$       \\
$N(1700)D_{13}$          &$\cdot\cdot\cdot$&$\cdot\cdot\cdot$&$\cdot\cdot\cdot$                                           &$150$&$1700$&$1.0$                               &$150$&$1700$&$1.0$     \\
$N(1675)D_{15}$        &$\cdot\cdot\cdot$&$\cdot\cdot\cdot$& $\cdot\cdot\cdot$                                            &$150$&$1675$&$1.0$                               &$150$&$1675$&$1.0$     \\
\hline\hline
\end{tabular}
\end{center}
\end{table*}

\begin{table*}
\begin{center}
\caption{Reduced $\chi^2$ per data point of the full model and that with one resonance or one background switched off obtained in a fit of the measured differential cross sections of the $\pi^+p\to \pi^+p$  and $\pi^-p\to \pi^-p,\pi^0 n$ reactions. Their corresponding $\chi^2s$ are labeled with $\chi^2_{\pi^+ p}$, $\chi^2_{\pi^- p}$ and $\chi^2_{\pi^0 n}$, respectively.}\label{consequence}
\begin{tabular}{ccccccccc}
\hline\hline
 ~~~&full model           ~~~&$n$ pole           ~~~&$\Delta(1232)P_{33}$  ~~~&$N(1535)S_{11}$  ~~~&$N(1520)D_{13}$ ~~~&$N(1440)P_{11}$ ~~~&$u$ channel ~~~&$t$ channel\\
\hline
$\chi^2_{\pi^+ p}$    ~~~&3.21   ~~~& $\cdot\cdot\cdot$        ~~~&74.84     ~~~&  $\cdot\cdot\cdot$        ~~~& $\cdot\cdot\cdot$        ~~~&       $\cdot\cdot\cdot$~~~&5.10   ~~~&7.71 \\
$\chi^2_{\pi^- p}$    ~~~&2.48    ~~~&9.89    ~~~&59.40     ~~~&3.29     ~~~&2.49    ~~~&4.55  ~~~&8.36    ~~~&4.33  \\
$\chi^2_{\pi^0 n}$    ~~~&3.60    ~~~&5.70     ~~~&147.21      ~~~&8.15      ~~~&3.87   ~~~&4.80   ~~~&19.14   ~~~&7.77  \\
\hline\hline
\end{tabular}
\end{center}
\end{table*}

\begin{figure*}[htbp]
\begin{center}
\centering \epsfxsize=16.8 cm \epsfbox{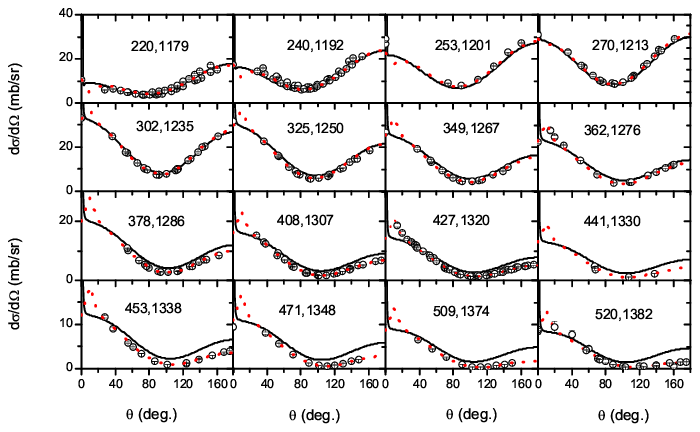} 
\vspace{-0.5cm} \caption{ Differential cross sections of the
$\pi^+ p \to \pi^+ p$ reaction compared with the experimental data (open circles) from~\cite{Pavan:2001gu,
Bussey:1973gz,Sadler:1987pi,Minehart:1981gi,Gordeev:1981ti} and the solutions (dotted curves) from the GWU group~\cite{INS:Data}. The first and second numbers in each figure correspond to the incoming $\pi$
momentum $P_{\pi}$ (MeV) and the $\pi N$ center-of-mass (c.m.) energy $W$ (MeV), respectively.}
\label{diff-1}
\end{center}
\end{figure*}

\begin{figure*}[htbp]
\begin{center}
\centering \epsfxsize=17.2 cm \epsfbox{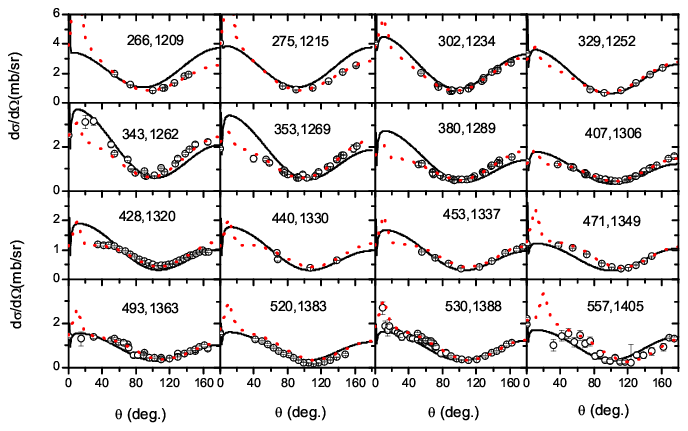} 
\vspace{-0.8cm} \caption{ Differential cross sections of the
$\pi^- p \to \pi^- p$ reaction compared with the experimental data (open circles) from~\cite{Bussey:1973gz,Pavan:2001gu,Minehart:1981gi,Gordeev:1981ti,Sadler:1987pi}
and the solutions (dotted curves) from the GWU group~\cite{INS:Data}.
The first and second numbers in each figure correspond to the incoming $\pi$
momentum $P_{\pi}$ (MeV) and the $\pi N$ c.m. energy $W$ (MeV), respectively.}
\label{diff-2}
\end{center}
\end{figure*}

\begin{figure*}[htbp]
\begin{center}
\centering \epsfxsize=17.2 cm \epsfbox{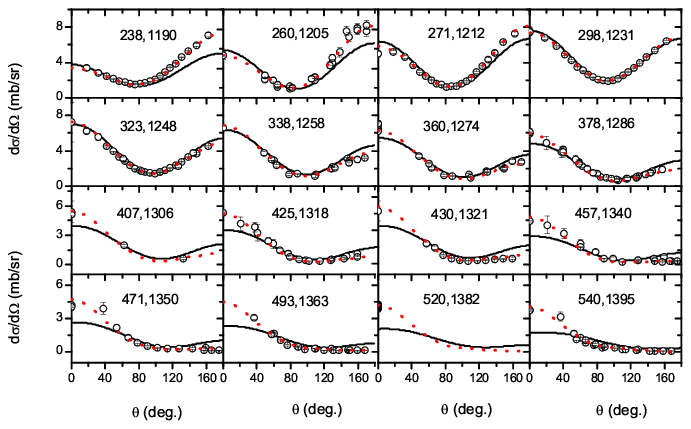} 
\vspace{-0.8cm} \caption{ Differential cross sections of the
reaction $\pi^- p \to \pi^0 n$ compared with the experimental data (open circles)
from~\cite{Jenefsky:1977br,Comiso:1974ee,Sadler:2004yq,Hauser:1971eh,Berardo:1972ay,Gaulard:1998sk} and the solutions (dotted curves) from the GWU group~\cite{INS:Data}. The first and second numbers in each figure correspond to the incoming $\pi$ momentum $P_{\pi}$ (MeV) and the $\pi N$ c.m. energy $W$ (MeV), respectively.} \label{diff-3}
\end{center}
\end{figure*}

\begin{figure*}[htbp]
\begin{center}
\centering \epsfxsize=16.8 cm \epsfbox{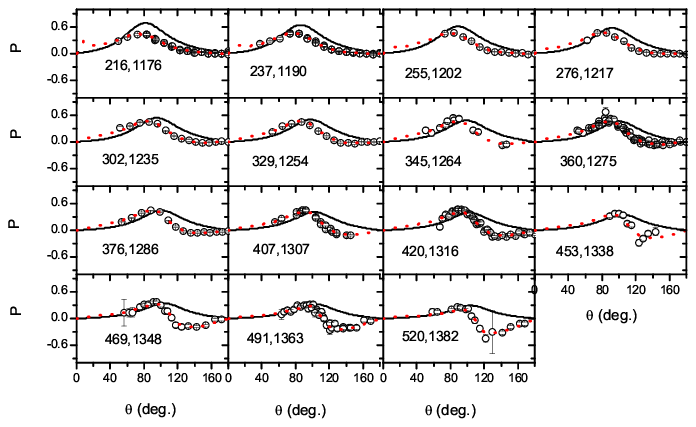} 
\vspace{-0.5cm} \caption{ Polarizations of the $\pi^+ p \to \pi^+ p$
reaction compared with experimental data (open circles) from~\cite{Hofman:1998gs,Amsler:1975hz,Sevior:1989nm,Bosshard:1991zp,Supek:1993qa,Dubal:1977ep,Abaev:1983pu,Mokhtari:1987iy} and the solutions (dotted curves) from the GWU group~\cite{INS:Data}.
 The first and second numbers in each figure correspond to the incoming $\pi$
momentum $P_{\pi}$ (MeV) and the $\pi N$ c.m. energy $W$ (MeV), respectively.}
\label{POL-1}
\end{center}
\end{figure*}

\begin{figure}[ht]
\centering \epsfxsize=7.8 cm \epsfbox{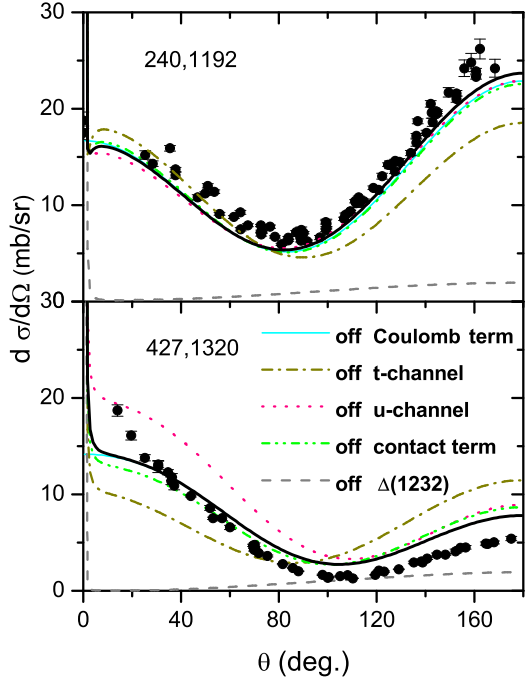} \vspace{-0.6cm}
\caption{ Effects of the main contributors on the differential cross sections of the reaction
$\pi^+p\rightarrow \pi^+p$ at two energy points $P_{\pi}$ =240, 427 MeV/c. The experimental data are taken from~\cite{Pavan:2001gu,Bussey:1973gz,Gordeev:1981ti}. The bold solid
curves correspond to the full model result. The predictions by
switching off the contributions from $\Delta(1232)P_{33}$, $u$- and
$t$-channel backgrounds are indicated explicitly by the legend in
the figures.} \label{Grapha1-1}
\end{figure}

\begin{figure}[htbp]
\begin{center}
\epsfxsize=7.4 cm \epsfbox{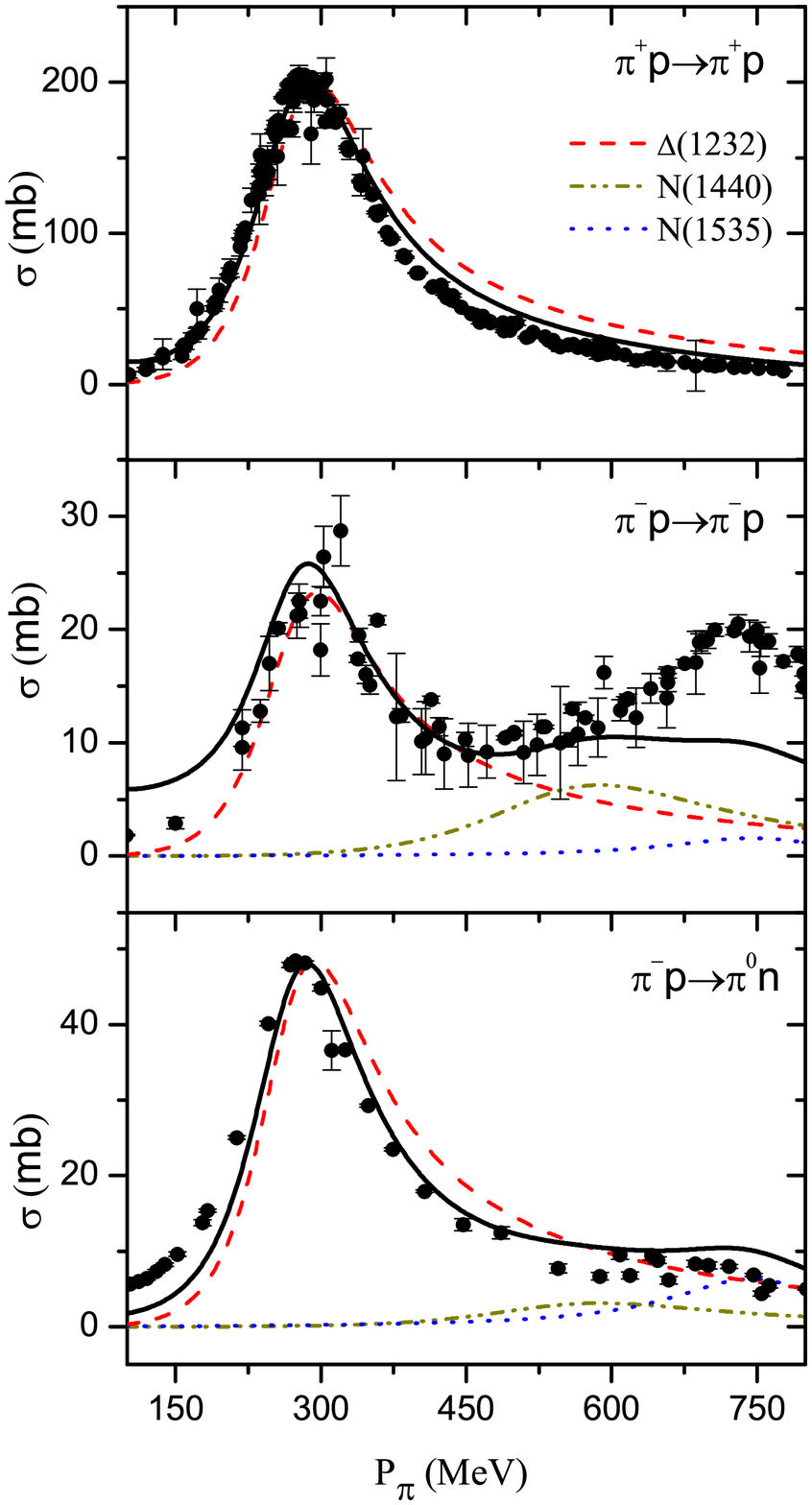} \vspace{-1.2cm} \caption{Predicted total cross sections of the reaction $\pi^+p \to \pi^+p$ and $\pi^- p \to \pi^- p,\pi^0 n$ compared with experimental data. The data for the $\pi^{\pm} p \to \pi^{\pm} p$ reactions are obtained from the PDG~\cite{PDG}, and the data of $\pi^- p \to \pi^0 n$ are taken from the Refs.~\cite{Sadler:2004yq,Comiso:1974ee,Breitschopf:2006gn,Chiu:1967zz,Bulos:1970zk,Berardo:1972ay}. The bold solid curves correspond to the full model result. In the figure, exclusive cross sections for $\Delta(1232)P_{33}$, $N(1535)S_{11}$, and $N(1440)P_{11}$ are indicated explicitly by the legends.  } \label{total2}
\end{center}
\end{figure}

\section{Calculation and analysis}

\subsection{Parameters}

In the calculation, we need four universal quark model parameters, i.e., the harmonic oscillator
parameter $\alpha$ and constituent quark masses for the $u$, $d$,
and $s$ quarks. They are well determined in our previous works, thus,
we fix them in our calculations. Their values are listed in Table~\ref{par1}.

In the $s$ and $u$ channels, the
quark-$\pi$-meson coupling is an overall parameter, which is related to the $\pi NN$ coupling via
the Goldberger-Treiman relation~\cite{Goldberger:1958tr}
\begin{eqnarray}
g_{\pi NN}=\frac{g_AM_N}{F_{\pi}},
\end{eqnarray}
where $g_A$ is the vector coupling for the $\pi$ mesons.
In the symmetrical quark model one can easily obtain $g_A=5/3$ and $5\sqrt{2}/6$
for charged and neutral pions, respectively. $F_{\pi}$ ($=f_{\pi}/\sqrt{2}\simeq 93$ MeV) is the pion-meson decay constant.
It should be remarked that the $\pi NN$ coupling is
a well-determined number:
\begin{eqnarray}
g_{\pi NN}=13.48,
\end{eqnarray}
thus we fix it in our calculations.

In the $t$ channel, there are two parameters, the coupling constants
$G_{V}a$ from the $\rho$-meson exchange and
$g_{SPP}g_{Sqq}$ from the $\sigma$-meson
exchange. We determine
these parameters by fitting the data, which are listed in Table~\ref{par1}.
It should be mentioned that the coupling constants
$G_{V}a$ and $g_{SPP}g_{Sqq}$ bear about a 30\% uncertainty.

In our framework, the $s$-channel resonance transition amplitude,
$\mathcal{O}_R$, is derived in the SU(6)$\otimes$O(3) symmetry limit.
In reality, the symmetry of SU(6)$\otimes$O(3) is
generally broken owing to some reasons.
To accommodate the symmetry-breaking and hadronic dressing effects,
following the idea of Ref.~\cite{Saghai:2001yd} we introduce a set coupling strength parameters, $C_R$, for each
resonance amplitude,
\begin{eqnarray}
\mathcal{O}_R\rightarrow C_R\mathcal{O}_R,
\end{eqnarray}
where $C_R$ should be determined by fitting the data. The deviations of $C_R$ from unity imply the
SU(6)$\otimes$O(3) symmetry breaking. The determined $C_R$ values are listed in
Table~\ref{compare}. It is found that  the $C_R$ parameters for
the $\Delta(1232)$ and $N(1440)$ resonances are notably larger than 1.
Thus, the SU(6)$\otimes$O(3) symmetry of these states is seriously broken by some effects,
which will be discussed later in details.

Furthermore, the masses and widths for the $s$-channel
resonances are important input parameters in the calculations.
For the main resonances $\Delta(1232)P_{33}$ and $N(1440)P_{11}$,
we vary their masses and widths in a proper range to better describe the data.
To be consistent with our previous study, we take the masses
and widths of $N(1535)S_{11}$ and $N(1520)D_{13}$ from~\cite{Xiao:2016dlf},
where the resonances parameters of $N(1535)S_{11}$ and $N(1520)D_{13}$
are well constrained.
The other resonances have few effects on the reactions, thus, their
masses and widths are taken from the PDG~\cite{PDG},
or the constituent quark model predictions~\cite{Isgur:1978xj,Isgur:1977ef,Isgur:1978wd}
if no experimental data are available. The masses and widths for some
low-lying resonances have been listed in
Table~\ref{compare}. It is found that the mass and width for
the $\Delta(1232)P_{33}$ are $M\simeq 1212$ MeV and $\Gamma\simeq 100$ MeV, respectively,
which are consistent with those extracted from the neutral
pion photoproduction processes in our previous work~\cite{Xiao:2015gra}.
It should be emphasized that our extracted mass and width for
the $\Delta(1232)P_{33}$ are quite close to the values of
the pole parametrization from the PDG~\cite{PDG}. The reason is that,
when we fit the data a constant resonance width $\Gamma_R$ is
used, which is similar to the pole parametrization.
Furthermore, we find that the $N(1440)P_{11}$ resonance seems to favour a
narrow width $\Gamma\simeq200$ MeV, which is also comparable to the values of
the pole parametrization from the PDG~\cite{PDG}.

In the $u$ channel, it is found that contributions from the $n\geq1$
shell resonances are negligibly small and insensitive to their
masses. Thus, the degenerate masses for the $n=1,2$ shell
resonances are taken in our calculations. The values have
been listed in Table~\ref{par1}.

Finally, it should be pointed out that all the adjustable parameters are
determined by globally fitting the measured differential cross
sections, which are obtained from~\cite{INS:Data}.
For the $\pi^+ p\rightarrow\pi^+
p$ reaction, we fit the differential cross sections
in the incoming pion-meson momentum range $P_{\pi}=260\sim 420$ MeV/c,
while for the $\pi^- p\rightarrow\pi^- p,\pi^0
n$ reactions, we fit the measured differential cross sections in the range
$P_{\pi}=260\sim 540$ MeV/c.
The data sets used in our fits are shown in
Figs.~\ref{diff-1},~\ref{diff-2} and ~\ref{diff-3}.
The reduced $\chi^2$s per data point obtained in our fits are
listed in Table~\ref{compare}. To clearly see the role of one
component in the reactions, the $\chi^2$s with one resonance or one
background switched off are also given in the Table~\ref{consequence}.

\subsection{$\pi^+p\rightarrow \pi^+p$}

The $\pi^+p\rightarrow \pi^+p$ process provides us a rather clear
channel to study the $\Delta$ resonances, because only the isospin $3/2$
resonances contribute here for the isospin selection rule. The
low-lying $\Delta$ resonances classified in the quark model are listed
in Table~\ref{amplitudes}. From the table we can see that in
a rather wide $\pi N$ center-of-mass (c.m.) energy range $W<1.6$ GeV,
only the $\Delta(1232)P_{33}$ resonance lies.
The higher resonances are the $S$-wave state $\Delta(1620)S_{31}$
and the $D$-wave state $\Delta(1700)D_{33}$, which may mainly contribute to
the reaction in the higher energy range $W>1.5$ GeV. Thus, the description
of the $\pi^+p\rightarrow \pi^+p$ reaction in the low energy region
becomes relatively simple.

The chiral quark model allows us study the
$\pi^+p\rightarrow \pi^+p$ reaction from the $\Delta(1232)$ resonance
region up to $W\simeq 1.4$ GeV.
Our fits of the differential cross
sections and polarizations compared with the data are shown in
Figs.~\ref{diff-1} and~\ref{POL-1},
respectively. From these figures, it is found that our fits are in a global agreement
with the experimental data in the c.m. energy range $W \simeq 1.2-1.4$
GeV, although in our calculations
the polarizations are overestimated slightly at the backward angles,
and the cross sections are overestimated slightly in the region $W> 1.3$ GeV.
New precise measurements with a good angle coverage are hoped to be carried
out in the future. For the limitations of the present model, our study cannot cover the higher
energy region $W> 1.4$ GeV.

To clearly understand the reaction mechanism of $\pi^+p\rightarrow \pi^+p$,
we show the main contributors one by one in Fig.~\ref{Grapha1-1}.
It is found that the interferences between the $\Delta(1232)P_{33}$ resonance and
the backgrounds of the $u$ and $t$ channels can roughly explain
the $\pi^+p\rightarrow \pi^+p$ reaction up to $W \simeq 1.4$
GeV. The Coulomb interactions may play an obvious role at the extremely forward angles.
Slight effects from the contact term can also be seen at the forward and backward angles.
The behavior of the contact term is similar to that of the $t$ channel.
No obvious effects of the higher resonances, such as
$\Delta(1620)S_{31}$ and $\Delta(1700)D_{33}$, are found in the low energy region
what we consider.

The cross sections around $W= 1.2$ GeV are
sensitive to the mass and width of $\Delta(1232)P_{33}$, which provide us
a good place to constrain the resonance parameters of $\Delta(1232)P_{33}$.
By fitting the measured total cross section with a momentum independent width (see Fig.~\ref{total2}), we obtain
that the mass and width of $\Delta(1232)P_{33}$ are $M\simeq 1212$ MeV and
$\Gamma\simeq 100$ MeV, respectively, with a uncertainty of several MeV. These
determined mass and width of $\Delta(1232)P_{33}$ are consistent with our
recent analysis of the pions photoproduction reactions~\cite{Xiao:2015gra},
and also are quite close to the values
of the pole parametrization from the PDG~\cite{PDG}.

Finally, it should be pointed out that to obtain a better description of the data, we should
enhance the $\Delta(1232)P_{33}$ contribution with a factor of $C_{\Delta}\simeq 3.0$,
which indicates that the $\Delta(1232) N \pi$ coupling may be underestimated
by a factor of $\sim \sqrt{3}$ in the SU(6)$\otimes$O(3) symmetry limit.
This underestimation is also found in the pions photoproduction reactions~\cite{Xiao:2015gra}
and the strong decays of $\Delta(1232)P_{33}$~\cite{Xiao:2013xi}.
The $\Delta(1232)P_{33}$ might not be a pure three-quark state~\cite{Sato:2000jf,
Bermuth:1988ms,Lu:1996rj,Faessler:2006ky,Aznauryan:2015zta,Sekihara:2015gvw},
some other contributions, such as meson-baryon component, may alter
the $\Delta(1232) N \pi$ coupling.

As a whole, from the $\Delta(1232)$ resonance region up to the $N(1440)$ resonance region,
the $\pi^+p$ elastic scattering can be reasonably understood
with the interferences between the $\Delta(1232)P_{33}$ resonance and
the backgrounds of the $u$ and $t$ channels. The extracted mass and width of $\Delta(1232)P_{33}$ are
quite close to the values of the pole parametrization~\cite{PDG}. The large
$\Delta(1232) N \pi$ coupling out of the quark model prediction may indicate that
$\Delta(1232)P_{33}$ may not be a pure three-quark state.

\subsection{$\pi^-p\rightarrow \pi^-p,\pi^0 n$}

Both the isospin-1/2 and -3/2 resonances contribute to the
$\pi^-p\rightarrow \pi^-p,\pi^0 n$ reactions.
From the spectrum classified in the quark model (see Table \ref{amplitudes}), we find that only the
$\Delta(1232)P_{33}$ and $N(1440)P_{11}$ resonances
lie within the $N(1440)P_{11}$ resonance region.
The higher resonances $N(1535,1650)S_{11}$, $\Delta(1620)S_{31}$,
$N(1520)D_{13}$ and $\Delta(1700)D_{33}$ are far from the $N(1440)P_{11}$ resonance region,
thus, their affects on these reactions should be small within this energy region. In this sense, the
$\pi^-p\rightarrow \pi^-p,\pi^0 n$ reactions might be good places
to study the properties of the $\Delta(1232)P_{33}$ and $N(1440)P_{11}$ resonances.

\begin{figure*}[htbp]
\begin{center}
\centering \epsfxsize=17.2 cm \epsfbox{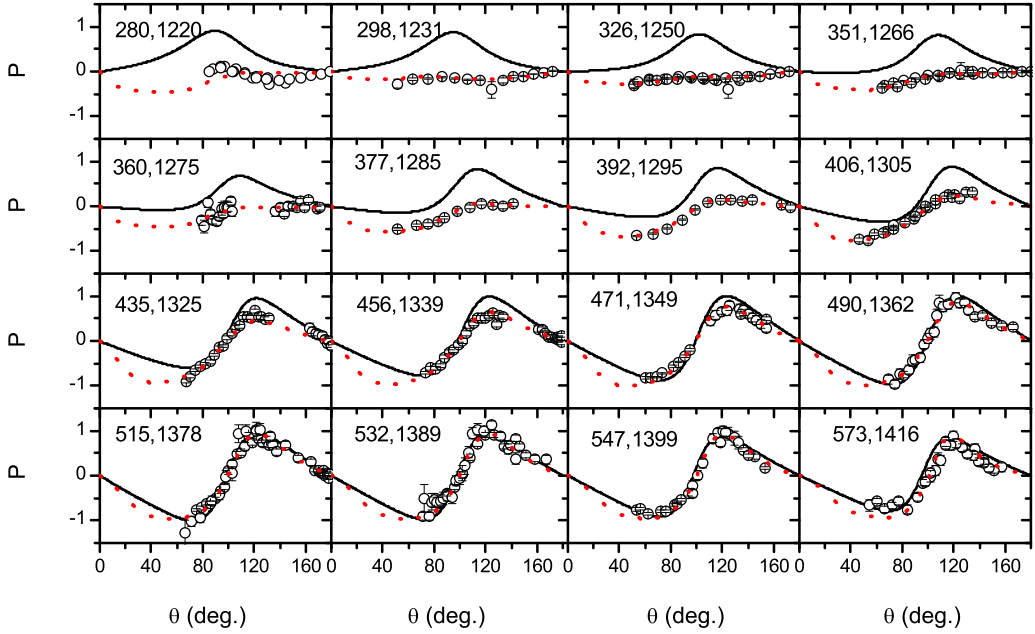} 
\vspace{-0.8cm} \caption{ Polarizations of the $\pi^- p
\to \pi^- p$ reaction compared with experimental
data (open circles) from~\cite{Hofman:1998gs,Sevior:1989nm,Arens:1968zz,Alder:1983my,Hofman:2003wt}
and the solutions (dotted curves) from the GWU group~\cite{INS:Data}.
The first and second numbers in each figure correspond to the incoming $\pi$
momentum $P_{\pi}$ (MeV) and the $\pi N$ c.m. energy $W$ (MeV), respectively.}
\label{POL-2}
\end{center}
\end{figure*}

\begin{figure*}[htbp]
\begin{center}
\centering \epsfxsize=17.2 cm \epsfbox{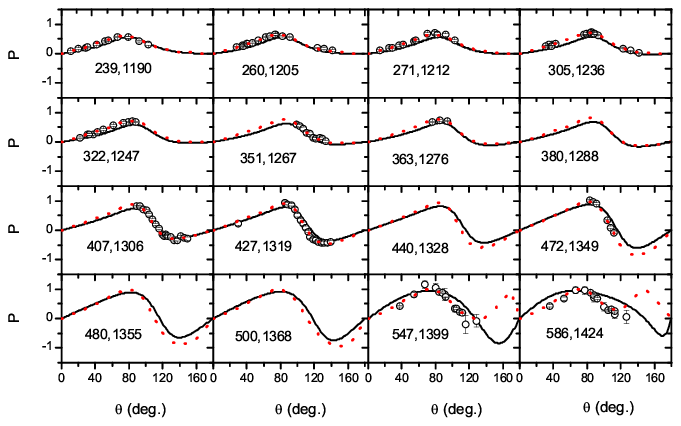} 
\vspace{-0.8cm} \caption{ Polarizations of the $\pi^- p
\to \pi^0 n$ reaction compared with the experimental
data (open circles) from~\cite{Gaulard:1998sk,Goergen:1990gp,Kim:1990de,Alder:1983my,Hill:1970xh}
and the solutions (dotted curves) from the GWU group~\cite{INS:Data}.
The first and second numbers in each figure correspond to the incoming $\pi$
momentum $P_{\pi}$ (MeV) and the $\pi N$ c.m. energy $W$ (MeV), respectively.}
\label{POL-3}
\end{center}
\end{figure*}

Based on our good understanding of $\pi^+ p\rightarrow\pi^+ p$,
we further study the $\pi^-p\rightarrow \pi^-p,\pi^0 n$
reactions. Our fits of the differential cross sections, total cross sections,
and polarizations compared with the data are
shown in Figs.~\ref{diff-2},~\ref{diff-3},~\ref{total2}-\ref{POL-3}.
From these figures, it is found that the experimental
data from $\Delta(1232)$ resonance region the up to the $N(1440)$ resonance region
are reasonably described within the chiral quark model. It should be mentioned
that there are remaining discrepancies in the polarizations of
$\pi^-p\rightarrow \pi^-p$ below $W=1.3$ GeV.
To gain more knowledge of these reactions, new precise measurements of the
polarizations with a good angle and energy coverage is hoped to be carried
out in the future.

\begin{figure*}[htbp]
\begin{center}
\centering \epsfxsize=7.4 cm \epsfbox{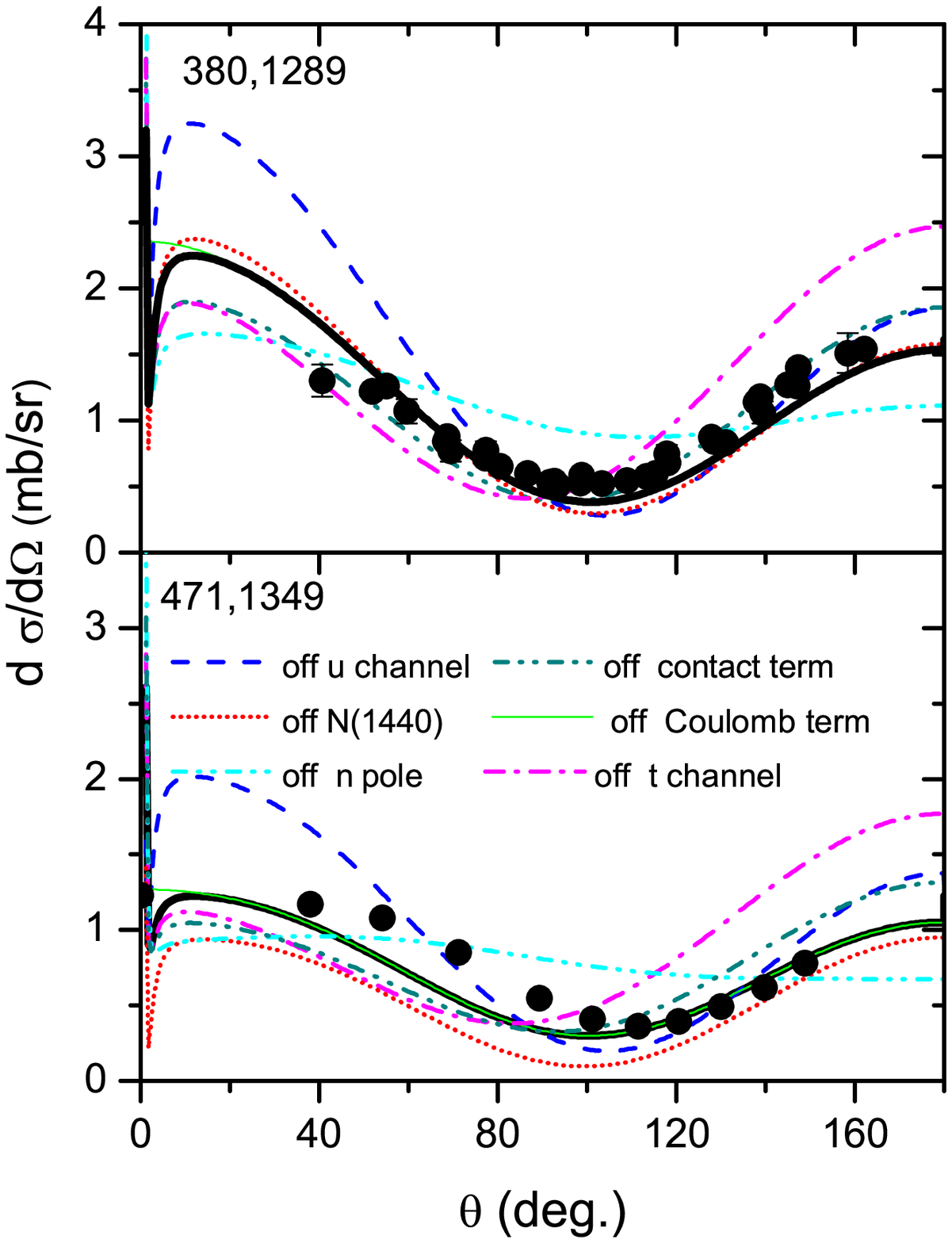} \epsfxsize=7.4 cm \epsfbox{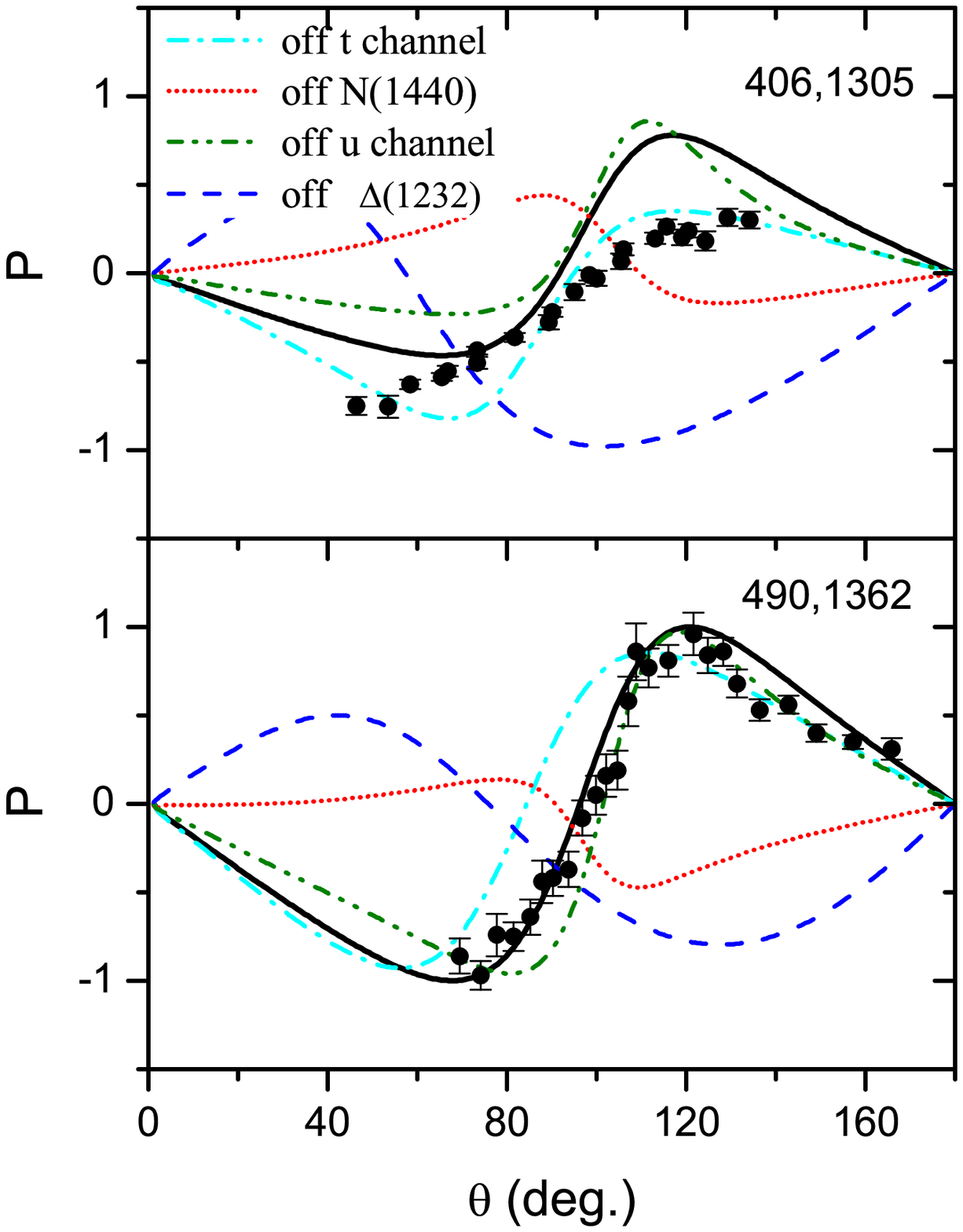} 
\vspace{-0.8 cm} \caption{Effects of the main contributors on the differential cross sections (left side) and polarizations (right side) of the
reaction $\pi^- p \to \pi^- p$. The experimental data are taken from
~\cite{Bussey:1973gz,Pavan:2001gu,Minehart:1981gi,Gordeev:1981ti,Sadler:1987pi,Sevior:1989nm}. The bold solid
curves correspond to the full model result. The results by
switching off the one of the main contributors in the resonances and non-resonance backgrounds are indicated explicitly by the legend
in the figures. The first and second numbers in each figure correspond to the incoming $\pi$
momentum $P_{\pi}$ (MeV) and the $\pi N$ c.m. energy $W$ (MeV), respectively.  } \label{DP-1}
\end{center}
\end{figure*}

\begin{figure*}[htbp]
\begin{center}
\epsfxsize=7.4 cm \epsfbox{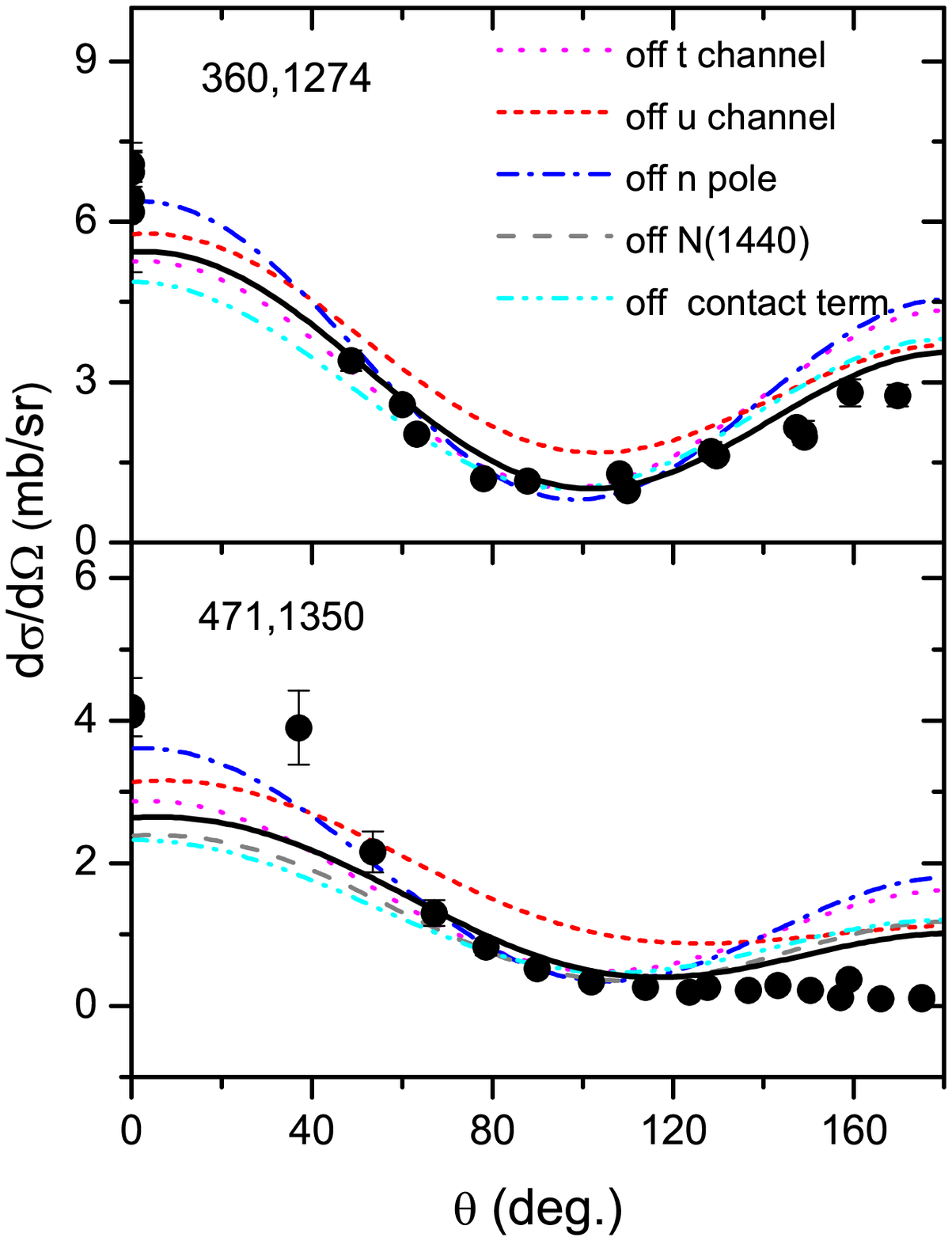} \epsfxsize=7.4 cm \epsfbox{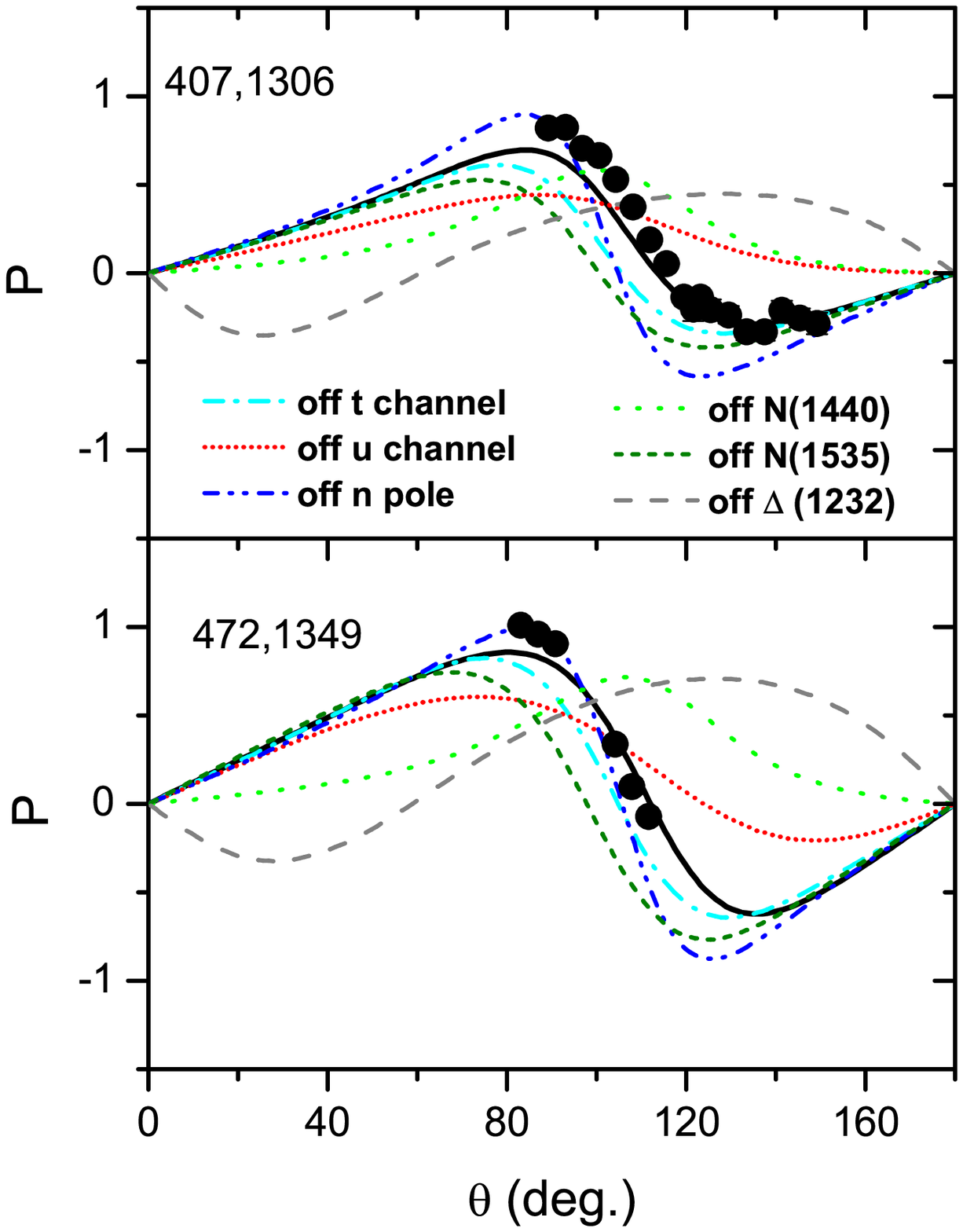} 
\vspace{-0.2cm} \caption{Effects of the main contributors on the differential cross sections (left side) and polarizations (right side) of the
reaction $\pi^- p \to \pi^0 n$. The experimental
data are taken from~\cite{Comiso:1974ee,Hauser:1971eh,Cheze:1974zd,Bayadilov:2004py,Lopatin:2002tr}. The bold solid
curves correspond to the full model result. The predictions by
switching off the one of the main contributors in the resonances and non-resonance backgrounds are indicated explicitly by the legend
in the figures. The first and second numbers in each figure correspond to the incoming $\pi$
momentum $P_{\pi}$ (MeV) and the $\pi N$ c.m. energy $W$ (MeV), respectively. } \label{DP-2}
\end{center}
\end{figure*}

To clearly understand the low energy reactions $\pi^-p\rightarrow \pi^-p,\pi^0 n$,
we show the main contributions to the
differential cross sections and polarizations in Figs.~\ref{DP-1} and~\ref{DP-2}, respectively.
From these figures, it is found that besides $\Delta(1232)P_{33}$ ,
the Roper $N(1440)P_{11}$ plays a crucial role in the
$\pi^-p\rightarrow \pi^-p,\pi^0 n$ reactions. Switching off their contributions,
we find that the differential cross sections and polarizations
have a notable change at both forward and backward angles.
It should be emphsized that a confirmed role of $N(1440)P_{11}$ can be more obviously seen from
the polarizations. Slight contributions from the $N(1535)S_{11}$ and $D(1520)D_{13}$ resonances
can extend to the $N(1440)$ resonance region as well, for simplicity, we do not show them in the figures.
No obvious effects of the higher resonances, such as
$\Delta(1620)S_{31}$, $\Delta(1700)D_{33}$, $D(1700)D_{13}$,
and $D(1675)D_{15}$ in the low energy regions.
The backgrounds from the $s$-channel nucleon pole, $u$ and $t$ channels
play important roles in the reactions.
The Coulomb interactions may play an obvious role at the extremely forward angles.
Slight effects from the contact term can also be seen at the forward and backward angles.

\begin{figure}[htbp]
\begin{center}
\centering \epsfxsize=7.0 cm \epsfbox{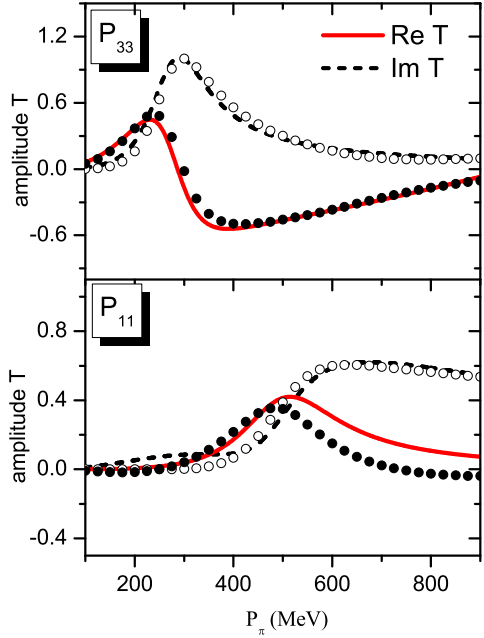} 
\vspace{-0.8 cm} \caption{The $\pi N$ partial amplitudes of $P_{33}$ and $P_{11}$. Solid (dashed) curves give the real (imaginary) parts of amplitudes extracted from our quark model. The filled (open) circles stand for the real (imaginary) parts (solution WI08~\cite{Arndt:2006bf} ) extracted by the GWU group~\cite{INS:Data}.} \label{DPp}
\end{center}
\end{figure}

Furthermore, to better understand the properties of
the $\Delta(1232)P_{33}$ and $N(1440)P_{11}$ resonances,
we also show our fits of the $P_{11}$ and $P_{33}$ amplitudes
to the solution WI08~\cite{Arndt:2006bf} from the GWU group~\cite{INS:Data} in fig.~\ref{DPp}.
Our results show a good agreement with the solution WI08.
Beyond the mass threshold of $N(1440)P_{11}$, although the
real part of the $P_{11}$ amplitude is overestimated in our quark model,
its tendency is similar to the solution WI08. It should be mentioned that in the
higher energy region $W\simeq 1.4$ GeV, our quark model begins to
lose its prediction ability, thus, our extracted properties
of $N(1440)P_{11}$ may be less reliable than those of $\Delta(1232)P_{33}$.

In the $\pi^-p\rightarrow \pi^-p,\pi^0 n$ reactions,
to obtain a good description of the data we also need enhance the
contribution of $\Delta(1232)P_{33}$ from the symmetric quark model
with a factor of $C_{\Delta}\simeq 3.0$. The extracted mass
and width of $\Delta(1232)P_{33}$ from these two reactions are
consistent with those from the $\pi^+p\to \pi^+p$ reaction.
At the mass threshold of $\Delta(1232)P_{33}$, our extracted ratios of the
total cross sections between these three $\pi N$ reactions
\begin{eqnarray}
&&\sigma (\pi^+p\to \pi^+p):\sigma (\pi^-p\to \pi^-p):\sigma (\pi^-p\to \pi^0n)\nonumber\\
&\simeq&205:25:48,
\end{eqnarray}
are quite close to the theoretical ratios $9:1:2$,
which indicates that the isospin symmetry well holds in these low energy $\pi N$ reactions.

Finally, it should be emphasized that confirmed roles of $N(1440)P_{11}$
have been seen in the $\pi^-p\rightarrow \pi^-p,\pi^0 n$ reactions,
which provide us a good opportunity to extract the properties of $N(1440)P_{11}$.
From the data of the $\pi^-p\rightarrow \pi^-p,\pi^0 n$ reactions,
we extract the mass and width of $N(1440)P_{11}$, $M\simeq 1400$ MeV and
$\Gamma\simeq200$ MeV, which
are close to those extracted from the pole parametrization~\cite{PDG}. Furthermore,
it should be mentioned that to well describe the data we need enhance the
contribution of the $N(1440)P_{11}$ from the symmetric quark model
with a rather large factor $C_{N(1440)}\simeq 23$, i.e.,
the $N(1440)N\pi$ coupling is a factor $\sim 4.8$ larger than
the prediction with the simple three-quark model,
which was also found by analyzing the strong decays of $N(1440)P_{11}$ in Refs.~\cite{JuliaDiaz:2004qr,Melde:2005hy}.
The unexpected large $N(1440)N\pi$ coupling indicates the exotic nature
of the $N(1440)P_{11}$ resonance. About the unusual properties of $N(1440)$,
there are many discussions in the literature~\cite{Kisslinger:1995yw,Krehl:1999km,Zou:2005xy,Zou:2010tc,Li:2006nm,JuliaDiaz:2006av,Liu:2016uzk,
Gegelia:2016xcw,Obukhovsky:2011sc,Obukhovsky:2013fpa,Yuan:2009st,Suzuki:2009nj,Lang:2016hnn}.

As a whole, besides the $\Delta(1232)P_{33}$ resonance, confirmed evidence of
the $N(1440)P_{11}$ resonance is found in the polarizations
of the $\pi^-p\rightarrow \pi^-p,\pi^0 n$ reactions.
The couplings of the $\Delta(1232) N \pi$ and $N(1440)N\pi$ predicted
from the simple three-quark model are about 1.7 and 4.8 times smaller than
the values extracted from the experimental data, respectively, which indicates that
the $\Delta(1232)P_{33}$ and $N(1440)P_{11}$ resonances cannot be pure three-quark states.
Finally, it should be mentioned that the $s$-channel nucleon pole, $u$- and
$t$-channel backgrounds play important roles in the reactions.

\section{Summary }

In this work, a combined study of the $\pi^+ p\rightarrow\pi^+ p$,
$\pi^-p\rightarrow \pi^- p$ and $\pi^- p \to \pi^0 n$ reactions have been
carried out within a chiral quark model. Our results show
a good global agreement with the data within the $N(1440)$ resonance region.

In these reactions, the resonance properties of $\Delta(1232)P_{33}$ are
constrained. By fitting the data with a momentum independent width, we obtain
the mass and width of $\Delta(1232)P_{33}$ are $M\simeq 1212$ MeV and
$\Gamma\simeq 100$ MeV, which are consistent with our
recent analysis of the pions photoproduction reactions~\cite{Xiao:2015gra}, and are quite close to
the values determined with the pole parametrization~\cite{PDG}.
The $\Delta(1232) N \pi$ coupling from the quark model is about
a factor of $\sim 1.7 $ smaller than that extracted from the data.
Some exotic components, such as the meson-baryon component, may alter
the $\Delta(1232) N \pi$ coupling, which should be studied further.

Confirmed roles of $N(1440)P_{11}$ resonance are found in both the
$\pi^-p\rightarrow \pi^- p$ and $\pi^- p \to \pi^0 n$ reactions.
The $N(1440)P_{11}$ has notable contributions to the
polarizations, although no obvious
effects can be seen in the total cross sections. The extracted mass
and width for $N(1440)P_{11}$ are $M\simeq 1400$ MeV and
$\Gamma\simeq 200$ MeV, respectively, which are close to
the values determined with the pole parametrization~\cite{PDG}.
The $N(1440)N\pi$ coupling extracted from the data is a factor of $\sim 4.8$ larger than
the symmetric quark model prediction. The unexpected large $N(1440)N\pi$ coupling suggests the uncommon properties
of the $N(1440)P_{11}$ resonance.

Starting from the incoming $\pi$-meson momentum $P_{\pi}\simeq 440$ MeV/c,
slight contributions of $N(1535)S_{11}$ and $N(1520)D_{13}$ are seen in both
the $\pi^-p\rightarrow \pi^- p$ and $\pi^- p \to \pi^0 n$
reactions. The backgrounds play remarkable roles in these three strong
interaction processes. The $t$- and $u$-channel backgrounds have notable
contributions to the $\pi^+ p\rightarrow\pi^+ p$ reactions.
While in the $\pi^-p\rightarrow \pi^- p,\pi^0 n$ reactions,
the $s$-channel nucleon pole and $t$- and $u$-channel backgrounds
play an important role.

Finally, it should be pointed out that with present model we cannot deliver higher accuracy descriptions
of the data because there are only a few adjustable parameters based on the SU(6)$\otimes$O(3)
symmetry, and all of the interactions are limited at the tree level. Furthermore, our present study is difficult to cover
the whole $N(1440)P_{11}$ resonance region, thus, the extracted
properties of $N(1440)P_{11}$ may bear a large uncertainty although
confirmed roles of $N(1440)P_{11}$ are found in the polarizations.
To uncover the uncommon nature of $N(1440)P_{11}$,
new precise measurements of the polarizations with a good angle and
energy coverage are expected to be carried out in the future.

\begin{table*}[htb]
\begin{center}
\caption{The $s$-channel resonance amplitudes within n=2 shell for the $\pi^+p\rightarrow \pi^+p$, $\pi^-p\rightarrow \pi^-p$ and $\pi^-p\rightarrow \pi^0n$ processes.
we have defined $M_{S}\equiv[\frac{\omega_{i}}{\mu_{q}}+|\mathbf{A}_{in}|\frac{2|\mathbf{k}|}{3\alpha^2}][\frac{\omega_{f}}{\mu_{q}}
+|\mathbf{A}_{out}|\frac{2|\mathbf{q}|}{3\alpha^2}]$, $M_{p}\equiv\frac{|A_{out}||A_{in}|}{|\mathbf{k}||\mathbf{q}|}$, $M_{P0}\equiv[\frac{\omega_{i}}{\mu_{q}}+|\mathbf{A}_{in}|\frac{|k|}{\alpha^2}][\frac{\omega_{f}}{\mu_{q}}
+|\mathbf{A}_{out}|\frac{|\mathbf{q}|}{\alpha^2}]$, $M_{D}\equiv|\mathbf{A}_{out}||\mathbf{A}_{in}|$, $M_{P2}\equiv[\frac{\omega_{i}}{\mu_{q}}
+|\mathbf{A}_{in}|\frac{2|\mathbf{k}|}{5\alpha^2}][\frac{\omega_{f}}{\mu_{q}}+|\mathbf{A}_{out}|
\frac{2|\mathbf{q}|}{5\alpha^2}]$, $M_{F}\equiv|\mathbf{A}_{out}||\mathbf{A}_{in}|\frac{ |\mathbf{k}||\mathbf{q}|}{9.0\alpha^4}$, $P'_{l}(z)\equiv\frac{\partial P_l(z)}{\partial z}$.
The functions $\mathbf{A}_{in}$ and $\mathbf{A}_{out}$ are defined by $\mathbf{A}_{in}\equiv-(\frac{\omega_{i}}{E_i+M_i}+1)\mathbf{k}$ and $\mathbf{A}_{out}\equiv-(\frac{\omega_{f}}{E_f+M_f}+1)\mathbf{q}$, respectively. The $\mu_{q}$
is a reduced mass at the quark model level, which equals to $1/\mu_{q}=2/m_{u}$ for $\pi N$ scattering processes.}\label{amplitudes}
\footnotesize
\begin{tabular}{|p{2.2cm}|p{2.0cm}|c|p{3.5cm}|p{3.5cm}|p{3.5cm}|}
\hline\hline
$resonance$         &$[N_6, ^{2S+1}N_3, n, l]$   &$O_{R}$ &~~~~$\pi^+p\rightarrow \pi^+p$       &~~~~$\pi^-p\rightarrow \pi^-p$      &~~~~$\pi^-p\rightarrow \pi^0n$  \\
\hline
$N(938)P_{11}$      &$|56,^28,0,0\rangle$   &$f(\theta)$ &~~~~$\cdot\cdot\cdot$&$\frac{25}{9}|\mathbf{A}_{in}||\mathbf{A}_{out}|P_{1}(z)$&$-\frac{25\sqrt{2}}{18}|\mathbf{A}_{in}||\mathbf{A}_{out}|P_{1}(z)$   \\
&&$g(\theta)$&~~~~$\cdot\cdot\cdot$&$\frac{25}{9}|\mathbf{A}_{in}||\mathbf{A}_{out}|\sin\theta P'_{1}(z)$&$-\frac{25\sqrt{2}}{18}|\mathbf{A}_{in}||\mathbf{A}_{out}|\sin\theta P'_{1}(z)$\\
\hline
$\Delta(1232)P_{33}$  &$|56,^410,0,0\rangle$  &$f(\theta)$ &$\frac{8}{3}|\mathbf{A}_{in}||\mathbf{A}_{out}|P_{1}(z)$&$\frac{8}{9}|\mathbf{A}_{in}||\mathbf{A}_{out}
|P_{1}(z)$&$\frac{8\sqrt{2}}{9}|\mathbf{A}_{in}||\mathbf{A}_{out}|P_{1}(z)$  \\
     &&$g(\theta)$&$-\frac{4}{3}|\mathbf{A}_{in}||\mathbf{A}_{out}|\sin\theta P'_{1}(z)$&$-\frac{4}{9}|\mathbf{A}_{in}||\mathbf{A}_{out}|\sin\theta P'_{1}(z)$&$-\frac{4\sqrt{2}}{9}|A_{in}||A_{out}|\sin\theta P'_{1}(z)$\\
\hline
$\Delta(1620)S_{31}$  &$|70,^210,1,1\rangle$  &$f(\theta)$ &$\frac{1}{24}M_{S}\alpha^2$&$\frac{1}{72}M_{S}\alpha^2$&$\frac{\sqrt{2}}{72}M_{S}\alpha^2$ \\
      &&$g(\theta)$&&$\cdot\cdot\cdot$&$\cdot\cdot\cdot$\\
\hline
$N(1535)S_{11}$       &$|70,^28,1,1\rangle$  &$f(\theta)$   &~~~~$\cdot\cdot\cdot$&$\frac{2}{9}M_{S}\alpha^2$&$-\frac{4\sqrt{2}}{36}M_{S}\alpha^2$ \\
  &&$g(\theta)$&~~~~$\cdot\cdot\cdot$&$\cdot\cdot\cdot$&$\cdot\cdot\cdot$\\
\hline
$N(1650)S_{11}$        &$|70,^48,1,1\rangle$  &$f(\theta)$   &~~~~$\cdot\cdot\cdot$&$\frac{1}{18}M_{S}\alpha^2$&$-\frac{\sqrt{2}}{36}M_{S}\alpha^2$ \\
   &&$g(\theta)$&~~~~$\cdot\cdot\cdot$&$\cdot\cdot\cdot$&$\cdot\cdot\cdot$\\
\hline
 $\Delta(1700)D_{33}$      &$|70,^210,1,1\rangle$&$f(\theta)$    &$\frac{1}{27}M_{D}\frac{|\mathbf{k}||\mathbf{q}|}{\alpha^2}P_{2}(z)$&$\frac{2}{81}M_{D}\frac{|\mathbf{k}||\mathbf{q}|}{\alpha^2}
 P_{2}(z)$&$\frac{2\sqrt{2}}{81}M_{D}\frac{|\mathbf{k}||\mathbf{q}|}{\alpha^2}P_{2}(z)$\\
   &&$g(\theta)$&$\frac{1}{54}M_{D}\frac{|\mathbf{k}||\mathbf{q}|}{\alpha^2}\sin\theta P'_{2}(z)$&$\frac{1}{81}M_{D}\frac{|\mathbf{k}||\mathbf{q}|}{\alpha^2}\sin\theta P'_{2}(z)$&$\frac{\sqrt{2}}{81}M_{D}\frac{|\mathbf{k}||\mathbf{q}|}{\alpha^2}\sin\theta P'_{2}(z)$\\
\hline
$N(1520)D_{13}$          &$|70,^28,1,1\rangle$   &$f(\theta)$     &~~~~$\cdot\cdot\cdot$&$\frac{40\times82}{41\times405}M_{D}\frac{|\mathbf{k}||\mathbf{q}|}{\alpha^2}P_{2}(z)$&$-\frac{40\times41\sqrt{2}}{41\times405}M_{D}\frac{|\mathbf{k}||\mathbf{q}|}{\alpha^2}P_{2}(z)$\\
  &&$g(\theta)$&~~~~$\cdot\cdot\cdot$&$\frac{40\times82}{41\times810}M_{D}\frac{|\mathbf{k}||\mathbf{q}|}{\alpha^2}\sin\theta P'_{2}(z)$&$-\frac{40\times41\sqrt{2}}{41\times810}M_{D}\frac{|\mathbf{k}||\mathbf{q}|}{\alpha^2}\sin\theta P'_{2}(z)$\\
\hline
$N(1700)D_{13}$        &$|70,^48,1,1\rangle$      &$f(\theta)$     &~~~~$\cdot\cdot\cdot$&$\frac{1\times82}{41\times405}M_{D}\frac{|\mathbf{k}||\mathbf{q}|}{\alpha^2}P_{2}(z)$&$-\frac{1\times41\sqrt{2}}{41\times405}M_{D}\frac{|\mathbf{k}||\mathbf{q}|}{\alpha^2}P_{2}(z)$\\
   &&$g(\theta)$&~~~~$\cdot\cdot\cdot$&$\frac{1\times82}{41\times810}M_{D}\frac{|\mathbf{k}||\mathbf{q}|}{\alpha^2}\sin\theta P'_{2}(z)$&$-\frac{1\times41\sqrt{2}}{41\times810}M_{D}\frac{|\mathbf{k}||\mathbf{q}|}{\alpha^2}\sin\theta P'_{2}(z)$\\
\hline
$N(1675)D_{I5}$     &$|70,^48,1,1\rangle$&$f(\theta)$    &~~~~$\cdot\cdot\cdot$&$\frac{2}{45}M_{D}\frac{|\mathbf{k}||\mathbf{q}|}{\alpha^2}P_{2}(z)$&$-\frac{\sqrt{2}}{45}M_{D}\frac{|\mathbf{k}||\mathbf{q}|}{\alpha^2}P_{2}(z)$\\
   &&$g(\theta)$&~~~~$\cdot\cdot\cdot$&$\frac{2}{135}M_{D}\frac{|\mathbf{k}||\mathbf{q}|}{\alpha^2}\sin\theta P'_{2}(z)$&$-\frac{\sqrt{2}}{135}M_{D}\frac{|\mathbf{k}||\mathbf{q}|}{\alpha^2}\sin\theta P'_{2}(z)$\\
\hline
$N(1440)P_{11}$    &$|56,^28,2,0\rangle$ &$f(\theta)$    &~~~~$\cdot\cdot\cdot$&$\frac{50\times11}{66\times81\times4}M_{P0}P_{1}(z)$&$-\frac{25\times11\sqrt{2}}{33\times162\times4}M_{P0}P_{1}(z)$ \\
  &&$g(\theta)$&~~~~$\cdot\cdot\cdot$&$\frac{50\times11}{66\times81\times4}M_{P0}\sin\theta P'_{1}(z)$&$-\frac{25\times11\sqrt{2}}{33\times162\times4}M_{P0}\sin\theta P'_{1}(z)$\\
\hline
$N(1710)P_{11}$      &$|70,^28,2,0\rangle$     &$f(\theta)$   &~~~~$\cdot\cdot\cdot$&$\frac{16\times11}{66\times81\times4}M_{P0}P_{1}(z)$&$-\frac{8\times11\sqrt{2}}{33\times162\times4}M_{P0}P_{1}(z)$\\
  &&$g(\theta)$&~~~~$\cdot\cdot\cdot$&$\frac{16\times11}{66\times81\times4}M_{P0}\sin\theta P'_{1}(z)$&$-\frac{8\times11\sqrt{2}}{33\times162\times4}M_{P0}\sin\theta P'_{1}(z)$\\
\hline
$\Delta(1900)P_{31}$       &$|70,^210,2,0\rangle$       &$f(\theta)$  &$\frac{1}{81\times8}M_{P0}P_{1}(z)$&$\frac{1}{81\times24}M_{P0}P_{1}(z)$&$\frac{\sqrt{2}}{81\times24}M_{P0}P_{1}(z)$\\
   &&$g(\theta)$&$\frac{1}{81\times8}M_{P0}\sin\theta P'_{1}(z)$&$\frac{1}{81\times24}M_{P0}\sin\theta P'_{1}(z)$&$\frac{\sqrt{2}}{81\times24}M_{P0}\sin\theta P'_{1}(z)$\\
\hline
$\Delta(1910)P_{31}$    &$|56,^410,2,2\rangle$          &$f(\theta)$   &$\frac{5}{81}M_{P2}|\mathbf{k}||\mathbf{q}|P_{1}(z)$&$\frac{5}{243}M_{P2}|\mathbf{k}||\mathbf{q}|P_{1}(z)$&$\frac{5\sqrt{2}}{243}M_{P2}|\mathbf{k}||\mathbf{q}|P_{1}(z)$\\
   &&$g(\theta)$&$\frac{5}{81}M_{P2}\sin\theta|\mathbf{k}||\mathbf{q}|P'_{1}(z)$&$\frac{5}{243}M_{P2}\sin\theta|\mathbf{k}||\mathbf{q}|P'_{1}(z)$&$\frac{5\sqrt{2}}{243}M_{P2}\sin\theta|\mathbf{k}||\mathbf{q}|P'_{1}(z)$\\
\hline
$N(1880)P_{11}$      &$|70,^48,2,2\rangle$      &$f(\theta)$   &~~~~$\cdot\cdot\cdot$&$\frac{5}{162\times6}M_{P2}|\mathbf{k}||\mathbf{q}|P_{1}(z)$&$-\frac{5\sqrt{2}}{162\times12}M_{P2}|\mathbf{k}||\mathbf{q}|P_{1}(z)$\\
  &&$g(\theta)$&~~~~$\cdot\cdot\cdot$&$\frac{5}{162\times6}M_{P2}\sin\theta|\mathbf{k}||\mathbf{q}|P'_{1}(z)$&$-\frac{5\sqrt{2}}{162\times12}M_{P2}\sin\theta|\mathbf{k}||\mathbf{q}|P'_{1}(z)$\\
\hline
$\Delta(1600)P_{33}$  &$|56,^410,2,0\rangle$  &$f(\theta)$   &$\frac{2}{81}M_{P0}|\mathbf{k}||\mathbf{q}|P_{1}(z)$&$\frac{2}{243}M_{P0}|\mathbf{k}||\mathbf{q}|P_{1}(z)$&$\frac{2\sqrt{2}}{243}M_{P0}|\mathbf{k}||\mathbf{q}|P_{1}(z)$\\
   &&$g(\theta)$&$-\frac{1}{81}M_{P0}\sin\theta|\mathbf{k}||\mathbf{q}|P'_{1}(z)$&$-\frac{1}{243}M_{P0}\sin\theta|\mathbf{k}||\mathbf{q}|P'_{1}(z)$&$-\frac{\sqrt{2}}{243}M_{P0}\sin\theta|\mathbf{k}||\mathbf{q}|P'_{1}(z)$\\
\hline
$N(?)P_{13}$       &$|70,^48,2,0\rangle$     &$f(\theta)$   &~~~~$\cdot\cdot\cdot$&$\frac{1}{162\times3}M_{P0}|\mathbf{k}||\mathbf{q}|P_{1}(z)$&$-\frac{\sqrt{2}}{162\times6}M_{P0}|\mathbf{k}||\mathbf{q}|P_{1}(z)$\\
  &&$g(\theta)$&~~~~$\cdot\cdot\cdot$&$-\frac{1}{162\times6}M_{P0}\sin\theta|\mathbf{k}||\mathbf{q}|P'_{1}(z)$&$\frac{\sqrt{2}}{162\times12}M_{P0}\sin\theta|\mathbf{k}||\mathbf{q}|P'_{1}(z)$\\
\hline
$N(1720)P_{13}$      &$|56,^28,2,2\rangle$    &$f(\theta)$     &~~~~$\cdot\cdot\cdot$&$\frac{25\times17}{34\times162\times3}M_{P2}|\mathbf{k}||\mathbf{q}|P_{1}(z)$&$-\frac{25\times5\times17\sqrt{2}}{34\times162\times12}M_{P2}|\mathbf{k}||\mathbf{q}|P_{1}(z)$\\
   &&$g(\theta)$&~~~~$\cdot\cdot\cdot$&$-\frac{25\times17}{34\times162\times6}M_{P2}\sin\theta|\mathbf{k}||\mathbf{q}|P'_{1}(z)$&$\frac{25\times5\times17\sqrt{2}}{34\times162\times12}M_{P2}\sin\theta|\mathbf{k}||\mathbf{q}|P'_{1}(z)$\\
\hline\hline
\end{tabular}
\end{center}
\end{table*}

\begin{table*}[htb]
\begin{center}
\caption{  A continuation of Table IV.} \label{amplitudes2}
\footnotesize
\begin{tabular}{|p{2.2cm}|p{2.0cm}|c|p{3.5cm}|p{3.5cm}|p{3.5cm}|}
\hline\hline
Resonance         &$[N_6, ^{2S+1}N_3, n, l]$   &$O_{R}$ &~~~~$\pi^+p\rightarrow \pi^+p$       &~~~~$\pi^-p\rightarrow \pi^-p$      &~~~~$\pi^-p\rightarrow \pi^0n$  \\
\hline
$\Delta(1920)P_{33}$      &$|56,^410,2,2\rangle$       &$f(\theta)$    &$\frac{5}{81}M_{P2}|\mathbf{k}||\mathbf{q}|P_{1}(z)$&$\frac{5}{81\times3}M_{P2}|\mathbf{k}||\mathbf{q}|P_{1}(z)$&$\frac{5\sqrt{2}}{81\times3}M_{P2}|\mathbf{k}||\mathbf{q}|P_{1}(z)$\\
   &&$g(\theta)$&$-\frac{5}{162}M_{P2}\sin\theta|\mathbf{k}||\mathbf{q}|P'_{1}(z)$&$-\frac{5}{162\times3}M_{P2}\sin\theta|\mathbf{k}||\mathbf{q}|P'_{1}(z)$&$-\frac{5\sqrt{2}}{162\times3}M_{P2}\sin\theta|\mathbf{k}||\mathbf{q}|P'_{1}(z)$\\
\hline
$N(1900)P_{13}$       &$|70,^28,2,2\rangle$     &$f(\theta)$   &~~~~$\cdot\cdot\cdot$&$\frac{8\times17}{34\times162\times3}M_{P2}|\mathbf{k}||\mathbf{q}|P_{1}(z)$&$-\frac{16\times5\times17\sqrt{2}}{34\times162\times12}M_{P2}|\mathbf{k}||\mathbf{q}|P_{1}(z)$\\
   &&$g(\theta)$&~~~~$\cdot\cdot\cdot$&$-\frac{8\times17}{34\times162\times6}M_{P2}\sin\theta|\mathbf{k}||\mathbf{q}|P'_{1}(z)$&$\frac{8\times5\times17\sqrt{2}}{34\times162\times12}M_{P2}\sin\theta|\mathbf{k}||\mathbf{q}|P'_{1}(z)$\\
\hline
$N(?)P_{13}$       &$|70,^48,2,2\rangle$     &$f(\theta)$    &~~~~$\cdot\cdot\cdot$&$\frac{1\times17}{34\times162\times3}M_{P2}|\mathbf{k}||\mathbf{q}|P_{1}(z)$&$-\frac{2\times5\times17\sqrt{2}}{34\times162\times12}M_{P2}|\mathbf{k}||\mathbf{q}|P_{1}(z)$\\
   &&$g(\theta)$&~~~~$\cdot\cdot\cdot$&$-\frac{1\times17}{34\times162\times6}M_{P2}\sin\theta|\mathbf{k}||\mathbf{q}|P'_{1}(z)$&$\frac{1\times5\times17\sqrt{2}}{34\times162\times12}M_{P2}\sin\theta|\mathbf{k}||\mathbf{q}|P'_{1}(z)$\\
\hline
$\Delta(?)P_{33}$       &$|70,^210,2,2\rangle$      &$f(\theta)$    &$\frac{5}{81\times8}M_{P2}|\mathbf{k}||\mathbf{q}|P_{1}(z)$&$\frac{5}{81\times24}M_{P2}|\mathbf{k}||\mathbf{q}|P_{1}(z)$&$\frac{5\sqrt{2}}{81\times24}M_{P2}|\mathbf{k}||\mathbf{q}|P_{1}(z)$\\
   &&$g(\theta)$&$-\frac{5}{162\times8}M_{P2}\sin\theta|\mathbf{k}||\mathbf{q}|P'_{1}(z)$&$-\frac{5}{162\times24}M_{P2}\sin\theta|\mathbf{k}||\mathbf{q}|P'_{1}(z)$&$-\frac{5\sqrt{2}}{162\times24}M_{P2}\sin\theta|\mathbf{k}||\mathbf{q}|P'_{1}(z)$\\
\hline
$N(1680)F_{15}$   &$|56,^28,2,2\rangle$   &$f(\theta)$   &~~~~$\cdot\cdot\cdot$&$\frac{175\times233}{233\times18\times35}M_{F}|\mathbf{k}||\mathbf{q}|P_{3}(z)$&$-\frac{175\times233\sqrt{2}}{233\times36\times35}M_{F}|\mathbf{k}||\mathbf{q}|P_{3}(z)$\\
   &&$g(\theta)$&~~~~$\cdot\cdot\cdot$&$\frac{175\times233}{233\times54\times35}M_{F}\sin\theta|\mathbf{k}||\mathbf{q}|P'_{3}(z)$&$-\frac{175\times233\sqrt{2}}{233\times108\times35}M_{F}\sin\theta|\mathbf{k}||\mathbf{q}|P'_{3}(z)$\\
\hline
$\Delta(1880)F_{35}$       &$|56,^410,2,2\rangle$      &$f(\theta)$    &$\frac{16\times23}{23\times12\times35}M_{F}|\mathbf{k}||\mathbf{q}|P_{3}(z)$&$\frac{16\times23}{23\times36\times35}M_{F}|\mathbf{k}||\mathbf{q}|P_{3}(z)$&$\frac{16\times23\sqrt{2}}{23\times36\times35}M_{F}|\mathbf{k}||\mathbf{q}|P_{3}(z)$\\
   &&$g(\theta)$&$\frac{16\times23}{23\times36\times35}M_{F}\sin\theta|\mathbf{k}||\mathbf{q}|P'_{3}(z)$&$\frac{16\times23}{23\times108\times35}M_{F}\sin\theta|\mathbf{k}||\mathbf{q}|P'_{3}(z)$&$\frac{16\times23\sqrt{2}}{23\times108\times35}M_{F}\sin\theta|\mathbf{k}||\mathbf{q}|P'_{3}(z)$\\
\hline
$N(1860)F_{15}$      &$|70,^28,2,2\rangle$       &$f(\theta)$   &~~~~$\cdot\cdot\cdot$&$\frac{56\times233}{233\times18\times35}M_{F}|\mathbf{k}||\mathbf{q}|P_{3}(z)$&$-\frac{56\times233\sqrt{2}}{233\times36\times35}M_{F}|\mathbf{k}||\mathbf{q}|P_{3}(z)$\\
  &&$g(\theta)$&~~~~$\cdot\cdot\cdot$&$\frac{56\times233}{233\times54\times35}M_{F}\sin\theta|\mathbf{k}||\mathbf{q}|P'_{3}(z)$&$-\frac{56\times233\sqrt{2}}{233\times108\times35}M_{F}\sin\theta|\mathbf{k}||\mathbf{q}|P'_{3}(z)$\\
\hline
$N(?)F_{15}$        &$|70,^48,2,2\rangle$      &$f(\theta)$   &~~~~$\cdot\cdot\cdot$&$\frac{2\times233}{233\times18\times35}M_{F}|\mathbf{k}||\mathbf{q}|P_{3}(z)$&$-\frac{2\times233\sqrt{2}}{233\times36\times35}M_{F}|\mathbf{k}||\mathbf{q}|P_{3}(z)$\\
 &&$g(\theta)$&~~~~$\cdot\cdot\cdot$&$\frac{2\times233}{233\times54\times35}M_{F}\sin\theta|\mathbf{k}||\mathbf{q}|P'_{3}(z)$&$-\frac{2\times233\sqrt{2}}{233\times108\times35}M_{F}\sin\theta|\mathbf{k}||\mathbf{q}|P'_{3}(z)$\\
\hline
$\Delta(2000)F_{35}$     &$|70,^210,2,2\rangle$       &$f(\theta)$  &$\frac{7\times23}{23\times12\times35}M_{F}|\mathbf{k}||\mathbf{q}|P_{3}(z)$&$\frac{7\times23}{23\times36\times35}M_{F}|\mathbf{k}||\mathbf{q}|P_{3}(z)$&$\frac{7\times23\sqrt{2}}{23\times36\times35}M_{F}|\mathbf{k}||\mathbf{q}|P_{3}(z)$\\
&&$g(\theta)$&$\frac{7\times23}{23\times36\times35}M_{F}\sin\theta|\mathbf{k}||\mathbf{q}|P'_{3}(z)$&$\frac{7\times23}{23\times108\times35}M_{F}\sin\theta|\mathbf{k}||\mathbf{q}|P'_{3}(z)$&$\frac{7\times23\sqrt{2}}{23\times108\times35}M_{F}\sin\theta|\mathbf{k}||\mathbf{q}|P'_{3}(z)$\\
\hline
$\Delta(?)F_{37}$  &$|56,^210,2,2\rangle$    &$f(\theta)$  &$\frac{8}{35}M_{F}|\mathbf{k}||\mathbf{q}|P_{3}(z)$&$\frac{8}{105}M_{F}|\mathbf{k}||\mathbf{q}|P_{3}(z)$&$\frac{8\sqrt{2}}{105}M_{F}|\mathbf{k}||\mathbf{q}|P_{3}(z)$\\
         &&$g(\theta)$&$-\frac{2}{35}M_{F}\sin\theta|\mathbf{k}||\mathbf{q}|P'_{3}(z)$&$-\frac{2}{105}M_{F}\sin\theta|\mathbf{k}||\mathbf{q}|P'_{3}(z)$&$-\frac{2\sqrt{2}}{105}M_{F}\sin\theta|\mathbf{k}||\mathbf{q}|P'_{3}(z)$\\
\hline
$N(?)F_{17}$
   &$|70,^48,2,2\rangle$&$f(\theta)$    &~~~~$\cdot\cdot\cdot$&$\frac{2}{105}M_{F}|\mathbf{k}||\mathbf{q}|P_{3}(z)$&$-\frac{\sqrt{2}}{210}M_{F}|\mathbf{k}||\mathbf{q}|P_{3}(z)$\\

        &&$g(\theta)$&~~~~$\cdot\cdot\cdot$&$-\frac{1}{210}M_{F}\sin\theta|\mathbf{k}||\mathbf{q}|P'_{3}(z)$&$\frac{\sqrt{2}}{420}M_{F}\sin\theta|\mathbf{k}||\mathbf{q}|P'_{3}(z)$\\
\hline\hline
\end{tabular}
\end{center}
\end{table*}

\section*{Acknowledgments}

We are grateful for useful discussions and suggestions from
Qiang Zhao, Fei Huang, Jia-Jun Wu, Ju-Jun Xie and Xu Cao.
This work is partly supported by the National Natural Science
Foundation of China (Grants No. 11075051 and No. 11375061), the
Hunan Provincial Natural Science Foundation (Grant No. 13JJ1018),
and the Hunan Provincial Innovation Foundation for Postgraduate.

\bibliographystyle{unsrt}

\end{document}